\documentclass	[prb,twocolumn,showpacs]{revtex4-1}

\usepackage{graphicx}
\usepackage{dcolumn}
\usepackage{bm}
\usepackage{amsmath}
\usepackage{epstopdf}
\usepackage{wasysym}
\usepackage{color}

\begin{document}
\title{Influence of the Dzyaloshinskii-Moriya interaction on the FMR spectrum of magnonic crystals and confined structures}
\author{M.~Mruczkiewicz$^{1}$ } \email{m.mru@amu.edu.pl}
\author{M.~Krawczyk$^{2}$ } \email{krawczyk@amu.edu.pl}
\affiliation{$^{1}$Institute of Electrical Engineering, Slovak Academy of Sciences, Dubravska cesta 9, 841 04 Bratislava, Slovakia\\
$^{2}$Faculty of Physics, Adam Mickiewicz University in Poznan, Umultowska 85, Pozna\'{n}, 61-614, Poland}

\date{\today}

\begin{abstract}

We study the effect of surface-induced Dzyaloshinskii-Moriya interaction (DMI) on the ferromagnetic resonance (FMR) spectrum 
of thickness-modulated one-dimensional magnonic crystals and isolated stripes. The DMI is found to substantially increases the intensity of absorption peaks and shifts the frequencies of the laterally quantized modes. The role of the DMI is determined by analyzing the amplitude and phase distributions of dynamic magnetic excitations calculated with frequency and time domain calculation methods. We propose experimentally realizable magnonic crystals and confined structures with multiple FMR absorption peaks. The frequency or magnetic field separation between FMR lines is exploited to propose method for estimation of the DMI strength.

\end{abstract}
\pacs{75.30.Ds, 75.40.Gb, 75.75.-c, 76.50.+g}

\maketitle

\section{Introduction}

Ultrathin structures with broken symmetry, such as multilayers Pt/Co/Ir or Pt/Co/Ta, are intensively studied
because of the Dzyaloshinskii-Moriya interaction (DMI) induced at the interfaces with the ferromagnetic metal. \cite{117,118} 
Surface-induced DMI leads to interesting phenomena, which include the occurrence of 
spiral magnetic states, skyrmion lattices or isolated skyrmions. 
The latter are the subject of extensive studies towards racetrack memory applications. \cite{zhang2015skyrmion} 
However, skyrmion configurations only occur with the DMI above a certain threshold level,
 are observed under low magnetic fields and at low temperatures. 
Thus, much effort is put in the material engineering to increase the DMI strength
and extend an area in the phase diagram in which nontrivial magnetic states can exist. 
One of the challenges in this area of research is to determine experimental values of the DMI, 
since the measurement of this interaction strength requires the use of a complex Brillouin light scattering (BLS) technique \cite{stashkevich2015,boulle2016room,belmeguenai2015interfacial}, time-resolved scanning Kerr microscopy \cite{korner2015interfacial} or electrically excited and detected spin wave (SW) transmission. \cite{lee2015all}

\begin{figure}[!ht]
\includegraphics[width=0.45\textwidth]{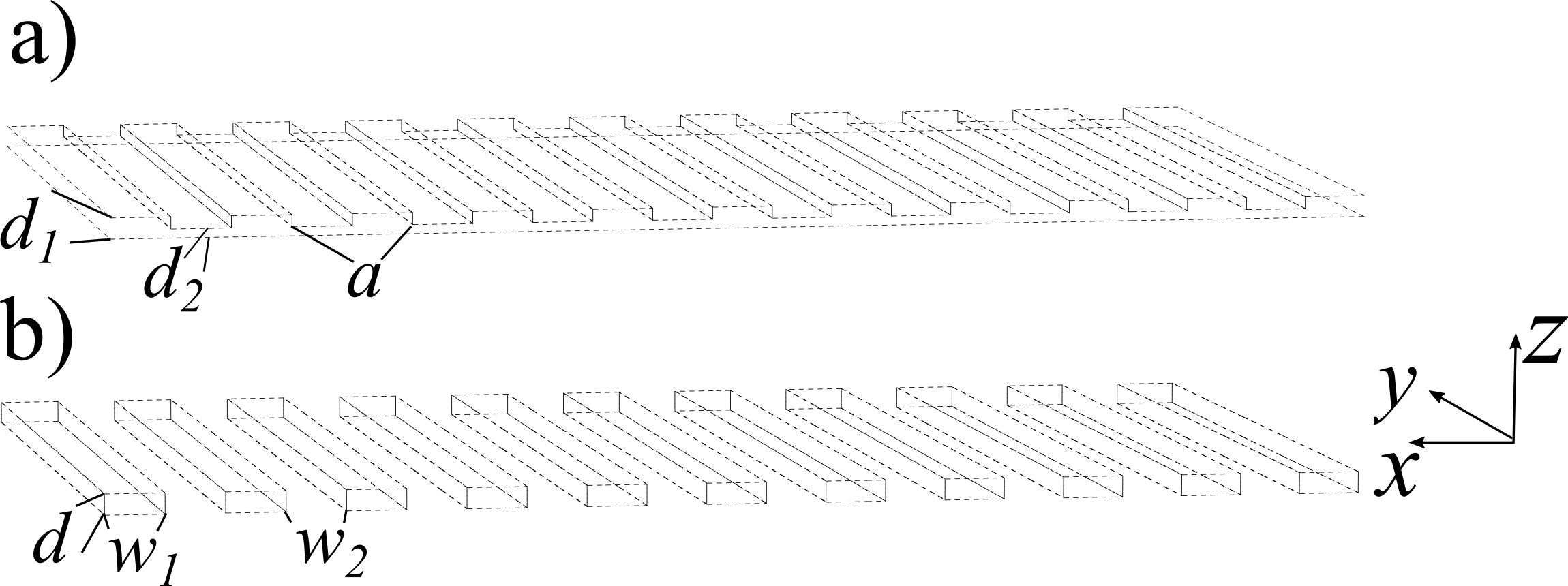}
\caption{(a) One-dimensional magnonic crystal consisting of a Co film with a periodically modulated thickness. 
(b) Array of non-interacting Co stripes.}
\label{structure}
\end{figure}

In the magnetically saturated state the DMI has a strong influence on the propagation of the SWs. 
It has been shown, the DMI results in an asymmetric dispersion relation, $f(-k_{1}) \neq f(k_{1})$, 
and nonreciprocal SW propagation.\cite{udvardi2009chiral,45,cortes2013influence} 
However, it has been demonstrated that the DMI has no influence on the ferromagnetic resonance (FMR) spectrum
in a uniform ferromagnetic film. \cite{cortes2013influence} 
Magnonic crystals (MCs) with DMI have been studied in terms of the SW propagation properties,\cite{ma2014interfacial} 
but the influence of the DMI on the FMR spectrum has not been reported to date. 
In this paper we focus on the effect of the DMI on the FMR spectrum of the one-dimensional (1D) MCs and magnetic stripes. 
We perform numerical calculations to demonstrate that the DMI increases the number and intensity of the absorption peaks 
 in both types of structures: MCs and isolated stripes. In both structures the DMI is found to split the peaks in the FMR spectrum.
On the basis of these findings we propose an experimental method for estimating the DMI strength. Taking into account the simplicity of the method, the broad accessibility of the experimental setup (cavity FMR or vector-network analyzer FMR) we expect, that the method can be used in studies of DMI materials in various emerging fields of physics, including magnonics, spintronics and also most recent spin-orbitronics. 
Moreover, the predicted increase in the FMR intensities of the high-frequency SWs might be of crucial importance for magnonic metamaterials with negative refractive index.\cite{Mikhaylovskiy2010,PhysRevB.86.024425}

The paper is organized as follows. In Section~II we present the model used in the calculations. 
In Section~III we study the dispersion relation in an MC with DMI, 
and demonstrate the effect of the DMI on SW modes with zero wavevector. 
The FMR spectra of MCs with DMI are presented in Section~IV. 
In Section~V we study the FMR spectrum and SW excitations in isolated stripes. 
Conclusions are presented in the closing Section~VI.

\section{Model}\label{Sec:Model}

We use the frequency-domain finite-element method~(FDFEM) 
and finite-difference time-domain~(FDTD) micromagnetic simulations 
to determine the influence of the DMI on the FMR spectrum of the studied structures.

The bias magnetic field $H_{0}$ is assumed to be strong enough to saturate the sample along the $y$~axis. 
In FDFEM, in the linear approximation, the magnetization vector, ${\bf M}$,  
can be represented as a sum of the static component, $(0, M_{\text{S}},0)$,  parallel to the $y$~axis ($M_{\text{S}}$ is the saturation magnetization), and the dynamic components lying in the ($x,z$)~plane, 
$\textbf{m}=(m_{x},0,m_{z})$. In the FDFEM model Maxwell's equations are considered in the magnetostatic approximation:
\begin{eqnarray}
\nabla \times  \textbf{h}({\bf r}) &=& \sigma \textbf{e}({\bf r}), \\ 
\nabla \times \textbf{e}({\bf r}) &=& -i \mu_{0} \omega ({\textbf h}({\bf r}) + {\textbf m}({\bf r})), \label{Eq:Max2}\\
\nabla \cdot {\bf B}&=&0,
\end{eqnarray}
where $\mu_{0}$ is the permeability of vacuum, ${\bf r}$ is the position vector, 
$\sigma$ is the conductivity of the ferromagnetic film, 
and $\omega$ is the angular frequency of magnetization oscillations; 
${\textbf e}$ is the electric field, 
${\textbf h}$ is  the dynamic magnetic field, 
and ${\textbf B}$ is the magnetic induction. 
In the considered geometry only the $y$~component of the electric field is related to the dynamic magnetic field, 
${\bf e} = (0,e_{y},0)$.\cite{Mru} 

Maxwell's equations are complemented with the damping-free Landau-Lifshitz equation of motion 
in a ferromagnetic film:
\begin{equation}
\frac{d   {\bf M}({\bf r},t)}{dt} = -\gamma \mu_{0}  {\bf M}({\bf r},t) \times {\bf H}_{\text{eff}}({\bf r},t),
\end{equation}
where  $\gamma$ is the gyromagnetic ratio and ${\bf H}_{\text{eff}}$ denotes the effective magnetic field: 
\begin{equation}
\begin{split}
{\bf H}_{\text{eff}}({\bf r},t)=H_{0} \hat{y} + \frac{1}{M_{\text{S}}} \nabla \cdot \left(\frac{2 A_{\text{ex}}}{\mu_{0} M_{\text{S}} }\right) \nabla {\bf M}({\bf r},t)+ \\
\frac{2 D}{\mu_{0} M_{\text{S}}^2}\left(\hat{y} \times \frac{\partial {\bf M}({\bf r},t)}{\partial x}\right) +  \textbf{h}({\bf r},t) \label{Eq:H_eff}
\end{split}
\end{equation}
where $A_{\text{ex}}$ is the exchange constant
and $D$ a parameter describing the strength of the DMI. 
In this model the DMI is considered in the form of an effective field. \cite{45} 

The solutions are assumed to have the form of a monochromatic Bloch wave:
 \begin{eqnarray}
{\bf m}({\bf r},t) &=& {\bf m'}({\bf r}) \exp (i \omega t) \exp (i k_{x} x),\\
{\bf h}({\bf r},t)&=& {\bf h'}({\bf r}) \exp (i \omega t) \exp (i k_{x} x),\\
{\bf e}({\bf r},t) &=& {\bf e'}({\bf r}) \exp (i \omega t) \exp (i k_{x} x),\label{Eq:Bloch}
 \end{eqnarray}
where  $k_{x}$ is a Bloch wavevector, $t$ is time, 
and the prime functions on the right side of the equations 
are periodic functions with a period equal to the lattice constant of the MC. Further, we define a unit cell with periodic boundary conditions on the external boundaries orthogonal to the $\widehat{x}$ and Dirichlet boundary conditions on the external boundaries orthogonal to the $\widehat{y}$, where all the functions are set to zero at distance of 1$\times10^{-5}$ m from the ferromagnetic sample. 

The above-described model is implemented in COMSOL Multiphysics®.\cite{Comsol} Nevertheless also an open source alternative can be used to solve this model.\footnote{For instance FEniCS\cite{logg2012automated} or Agros2D\cite{karban2013numerical}. In particular, the functionality of FEniCS allows to define arbitrary equations, periodic boundary conditions and use of the eigenvalue solver, necessary for solving problem described in this manuscript. In addition, stated problem could be solved also with spectral methods, such as plane wave method implemented for SWs.\cite{krawczyk2008plane}} In FDFEM we have used the triangular discretization with maximum element size 0.4 nm inside the magnetic material and 40 nm outside. The element growth rate was chosen to 1.1, to ensure the sufficiently small elements near the magnetic material.

The assumption concerning the magnetic ground state is only valid 
in a certain range of magnetic field and DMI strength parameter.
A DMI threshold $D_{\text{th}}$ is set at the level where the ground state of the structure 
will cease to be a single-domain collinear alignment. 
According to Ref.~[\onlinecite{45}] $D_{\text{th}}$ can be estimated by the equation:
\begin{equation}
D_{\text{th}}=\sqrt{\mu_{0} M_{\text{S}} A_{\text{ex}}\left[  H_{0}+\frac{M_{\text{S}}}{2}+\sqrt{  H_{0}\left(  H_{0}+M_{\text{S}}\right)}\right]}. \label{Eq:treshold}
 \end{equation}
This gives an approximate threshold value of 4.6 mJ/m$^2$ for $\mu_{0}  H_{0}=0.1$~T and 3.5 mJ/m$^2$ for $\mu_{0}  H_{0}=0$~T. 

The FDTD simulations are performed using mumax$^3$,\cite{4899186} 
with implemented surface-induced DMI and no assumptions concerning the magnetization alignment. Nevertheless, in the selected DMI parameter strength between $D=0$---2 mJ/m$^2$ the saturated state is preserved. The discretization cell used in FDTD was $2.5$ nm $\times$ $1$ nm $\times$ $1$ nm for the MCs and $1$ nm $\times$ $1$ nm $\times$ $1$ nm for the stripe. 

\begin{figure}[!ht]
\includegraphics[width=0.45\textwidth]{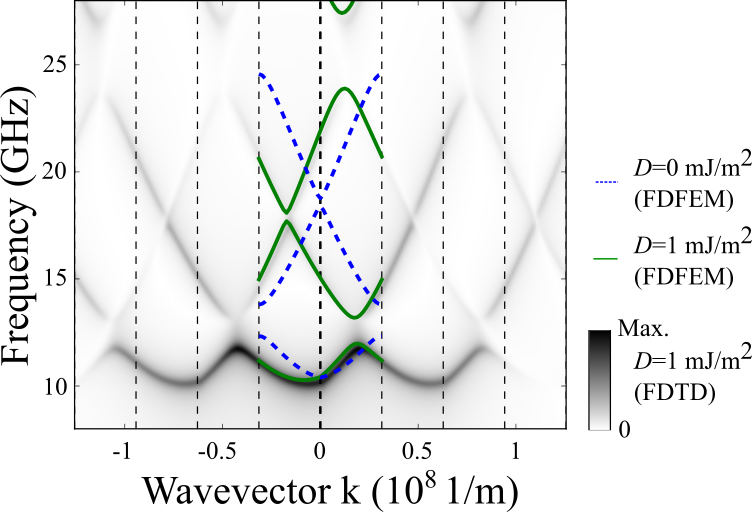}
\caption{(Color online) Dispersion relation of an MC consisting of a Co film with modulated thickness; 
$M_{S}$~=~0.956$\times10^6$ A/m, $A_{\text{ex}}$~=~2.1$\times10^{-11}$ J/m, $\mu_0 H_0$~=~100 mT. 
The black and white color map represents the results of FDTD simulations for $D=1$ mJ/m$^2$. 
Plotted on top are FDFEM calculation results 
for $D=0$ mJ/m$^2$ (blue dashed line) and $D=1$ mJ/m$^2$ (green solid line).}
\label{disp}
\end{figure}

%

Throughout the paper we use the same material parameters: 
saturation magnetization $M_{\text{S}}$ = 0.956 $\times 10^6$ A/m, exchange constant $A_{\text{ex}}$~=~2.1$\times10^{-11}$ J/m,
DMI strength parameter $D=0$---2 mJ/m$^2$, 
and out-of-plane magnetic anisotropy $K_{u}$=0---0.17 $\times 10^6$ J/m$^3$. 
This set of parameters is comparable to those measured in the Pt/Co/Ir structure.\cite{moreau2016additive} 
The assumed value of the damping parameter taken into account in FDTD is $\alpha=0.01$, 
is characteristic of an ultrathin Co film where FMR measurements were performed using a coplanar waveguide.\cite{beaujour2006magnetization,berger2014magnetization} 

\section{SW dispersion and profile characteristics in MC\lowercase{s} with DMI}

Figure~\ref{structure}(a) presents the investigated MC, 
consisting of an ultrathin Co layer with a periodically modulated thickness. 
The alternating regions of thickness $d_{1}=1$~nm and $d_{2}=0.5$~nm have an equal width of 50 nm, 
and the periodicity of the MC is $a=100$~nm. $K_{u}$ is set to 0 in the 
investigation of SWs in MCs below.

\begin{figure}[!ht]
\includegraphics[width=0.45\textwidth]{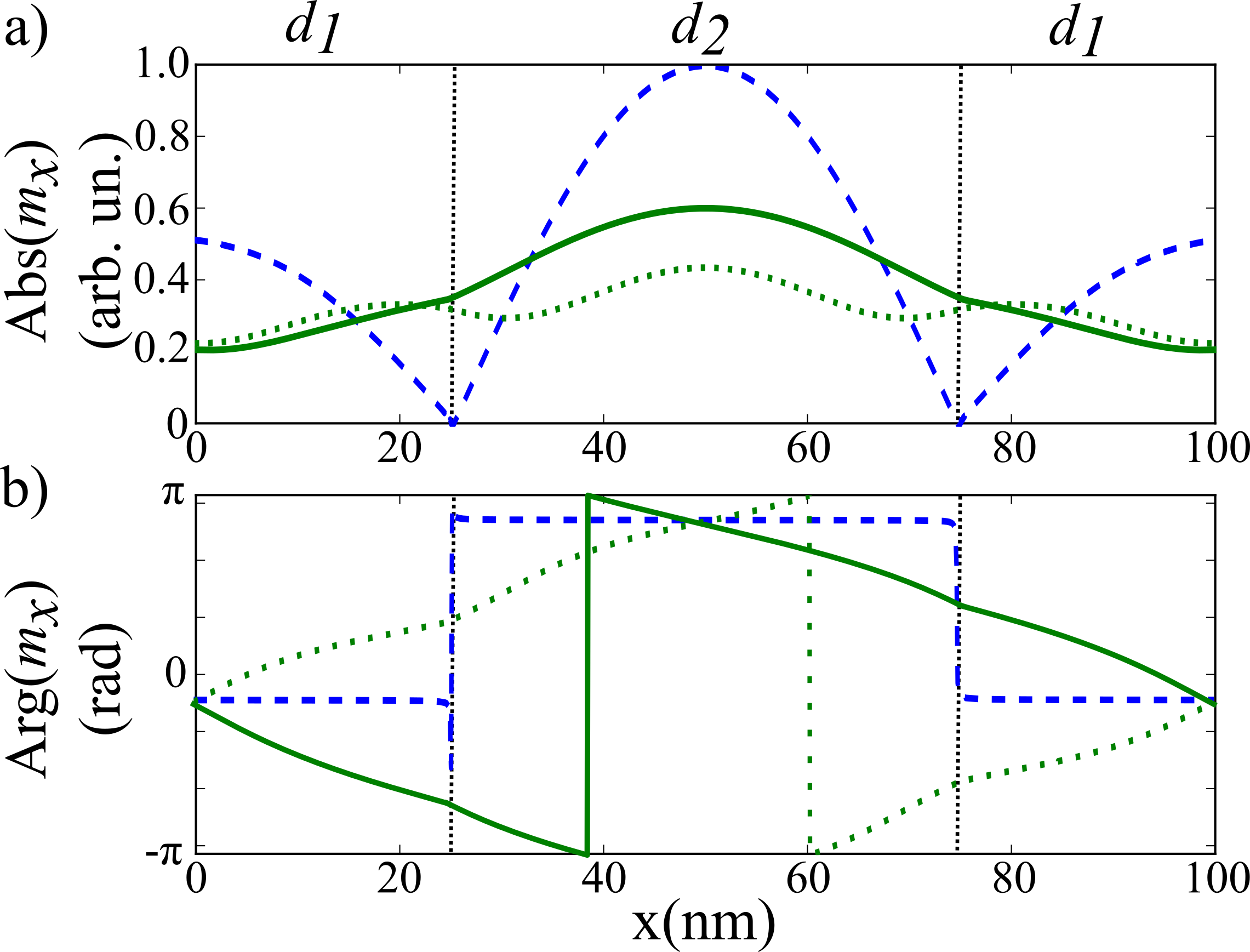}
\caption{(Color online) (a) Amplitude and (b) phase profiles of SW modes: 
$k_x=2 \pi /a$ for $D=0$~mJ/m$^2$ (blue dashed line, mode~ii in Fig.~\ref{FMR_MC}), 
$k_x=-2 \pi / a$ for $D=1$~mJ/m$^2$ (green continuous line, mode~i in Fig.~\ref{FMR_MC}),
and $k_x=2 \pi / a$  for $D=1$~mJ/m$^2$ (green dotted line, mode~iii in Fig.~\ref{FMR_MC}). }
\label{profiles}
\end{figure}

Figure~\ref{disp} presents the dispersion relations in the MC with a periodically modulated thickness 
with and without the DMI: 
for $D=0$ mJ/m$^2$ (blue dashed line, FDFEM results) 
and $D=1$ mJ/m$^2$ (green continuous line FDFEM and black-and-white color map, FDTD results). 
The results of the frequency-domain calculations and micromagnetic simulations are in good agreement. 
The FDFEM results are plotted only within the 1st Brillouin zone~(BZ), 
but the solutions repeat periodically 
with a period equal to the reciprocal lattice vector $G=\frac{2 \pi}{a} \approx 0.63 \times 10^8$ m$^{-1}$, according with the Bloch theorem Eq.~(\ref{Eq:Bloch}). 

In the MC without DMI the dispersion is reciprocal; the periodicity of the structure results in the occurrence of 
magnonic band gaps at the wavevectors fulfilling the Bragg condition,\cite{beginin2012bragg} 
i.e., for $k_x = n \pi /a$, where $n$ is an integer.
In this study we focus only on modes with wavevectors $k_x=2 n \pi /a$, 
which can be observed in FMR measurements,\cite{Mruczkiewicz13, PhysRevB.86.024425} 
their wavelength being an integer multiple of the lattice constant.  
Since the Bragg condition is always fulfilled for and the band gaps always open at $k_x=n \pi /a$, 
in the MC without DMI the modes have a zero group velocity, $V_{\text{gr}} = 0$, and are standing waves,\cite{Mruczkiewicz13} 
as indicated by the profile 
of the absolute value of the amplitude of the $x$~component of the dynamic magnetization for~$k_x=2 \pi /a$, 
shown in Fig.~\ref{profiles}(a) (blue dashed line). 
Nodes are found at the interfaces between MC segments with thickness $d_{1}$ and $d_{2}$.
Fig.~\ref{profiles}(b) shows the phase profile of the $x$ component of the dynamic magnetization. 
The phase is constant within each segment, but differs by $\pi$ between adjacent segments, which means antiphase oscillations. 
Since the asymmetry between segments in the amplitude distribution is slight, 
this mode can be excited, though with a low intensity, by a uniform external microwave magnetic field.

\begin{figure}[!ht]
\includegraphics[width=0.45\textwidth]{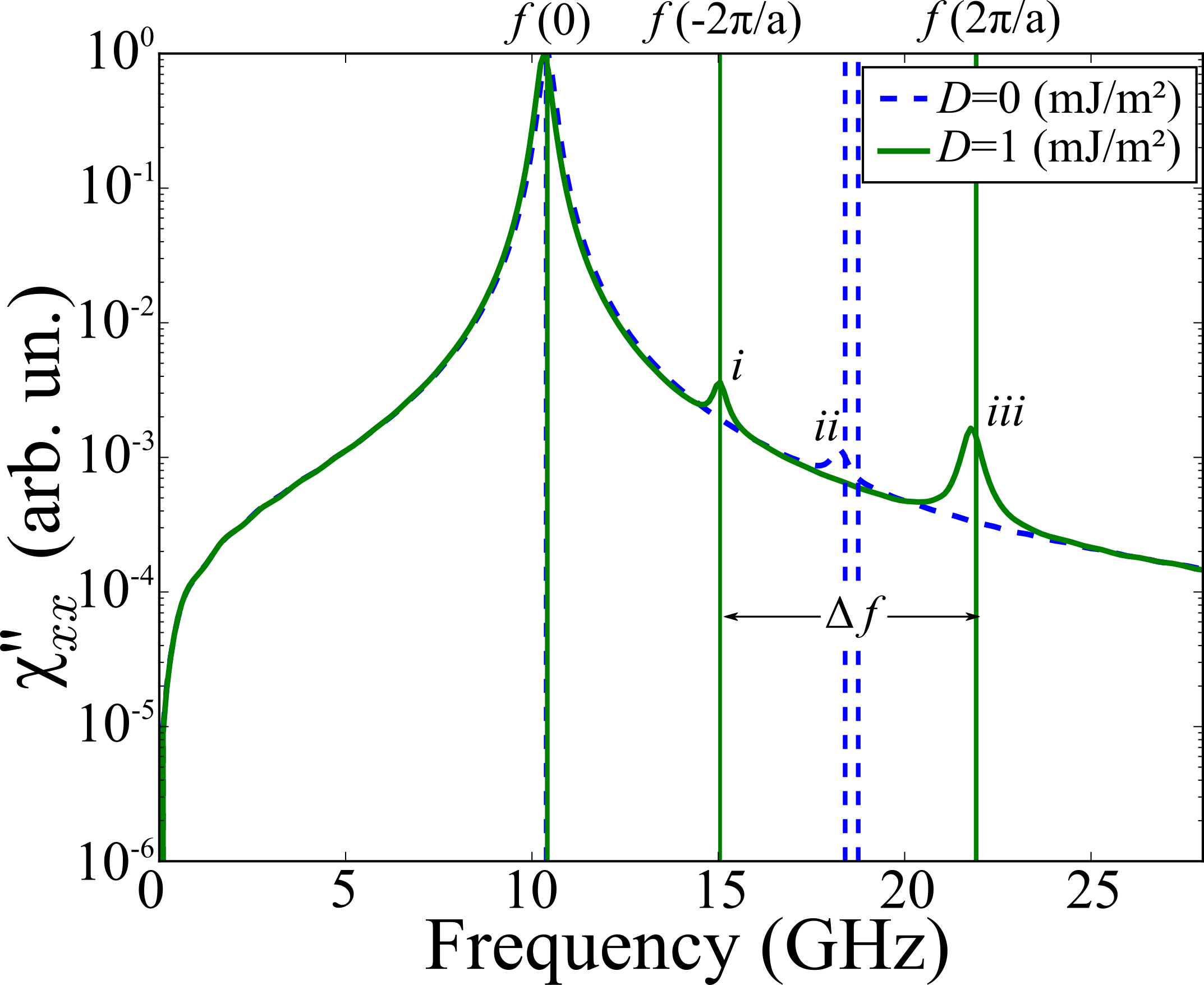}
\caption{(Color online) FMR frequency spectrum of the studied MC (Co film with modulated thickness); $\mu_0 H_{0}$~=~100~mT. 
Results of FDTD simulations for $D=0$ mJ/m$^2$ (blue dashed line) and $D=1$ mJ/m$^2$ (green solid line). The vertical lines indicate the eigenmode resonance position calculated with FDFEM. Due to the symmetry of the modes in the structure with $D=0$ 
only one mode around 18 GHz is observed in FDTD simulations.}
\label{FMR_MC}
\end{figure}
A different picture is obtained in the structure with the DMI. 
In a homogeneous ferromagnetic film the DMI results in a nonreciprocal dispersion relation, 
since it introduces a term proportional to~$k_x$:\cite{45, di2015asymmetric, stashkevich2015}
\begin{widetext}
 \begin{equation}
f=\frac{1}{2 \pi} \sqrt{(\gamma \mu_{0} H_{0}+ \omega_{\text{ex}} k_{x}^2)(\gamma \mu_{0} H_{0}+ \omega_{\text{ex}} k_{x}^2+\omega_{M})+\frac{\omega_{M}^2}{4} (1 - e^{-2 |k_x| d})} + \frac{\gamma D k_x}{\pi M_{S}},
\label{eq_1}
 \end{equation}
\end{widetext}
where $\omega_{M} = \gamma \mu_0 M_{S}$, $\omega_{\text{ex}} = \frac{2A_{\text{ex}}}{\mu_0 M_{S}}$, and $d$ is the film thickness.
Consequently, also in an MC with the DMI the dispersion relation is nonreciprocal (Fig.~\ref{disp}, green solid line). 
The exchange Bragg condition is fulfilled for wavevectors 
that are not integer multiples of $\pi / a$,\cite{yeh1979electromagnetic,Mruczkiewicz.2013b} 
and the band gaps are shifted away from the boundary and center of the BZ.\cite{Mruczkiewicz.2013b, lisenkov2015nonreciprocity, ma2014interfacial}

The dispersion in the MC with the DMI indicates 
that modes with wavevectors $k_x=2 n \pi /a$ are not standing waves with $V_{\text{gr}} = 0$, 
but propagating waves with a nonzero
group velocity, $V_{\text{gr}} \neq 0 $. 
This results in their significantly modified profile. 
Fig.~\ref{profiles}(a) shows the amplitude profiles of the mode 
in the 2nd band with $k_x = -2 \pi /a$ (green solid line)  
and with $k_x = 2 \pi /a$ (green dotted line) for $D=1$ mJ/m$^2$. 
The main difference with respect to the profile observed in the MC without DMI is the lack of nodes. 
Presented in Fig.~\ref{profiles}(b) the phase profiles of these SWs (green solid and dotted lines, respectively) 
show a continuous change of the phase along the $x$~direction (by $2\pi$ along a unit cell). 
Since an FMR absorption peak is proportional to $P \propto \left| \int \text{Abs}({\bf m}) e^{-i \text{Arg}({\bf m})} dr \right|^2 $, \cite{bihler2009spin}
modes with modified phase and amplitude in an MC with a nonreciprocal dispersion 
are expected to be observed with a relatively high intensity in an FMR experiment.

\section{FMR spectra of MCs}

In order to show the qualitative influence 
of the nonreciprocal dispersion and the consequent modification of the SW profiles on the FMR spectrum 
we have performed time-domain micromagnetic simulations with uniform microwave magnetic field excitation. 
By comparing the Fourier transform of the $x$~component of the magnetization 
with the $x$~component of the microwave exciting magnetic field
we have obtained the frequency dependence of the imaginary part of the susceptibility tensor, $\chi_{xx}^{\prime \prime}(f)$. \cite{gerardin2000micromagnetics}

Plotted in Fig.~\ref{FMR_MC}, the $\chi_{xx}^{\prime \prime}(f)$ dependencies 
obtained for $D=0$~mJ/m$^2$ (blue dashed line) and $D$~=~1~mJ/m$^2$ (green solid line) 
represent the FMR spectra of the MC with and without the DMI, respectively.
The DMI is found to have a significant impact on the high-frequency part of the FMR spectrum. 
The frequency of the fundamental excitation remains unchanged, though. 
The peaks and modes presented in Fig.~\ref{profiles} are labeled $i$, $ii$ and $iii$ in Fig.~\ref{FMR_MC}. 
Due to the symmetry of the modes in the structure with $D=0$ 
only one mode around 18 GHz is observed in the corresponding spectrum (see the Fig. 
\ref{FMR_MC}, blue dashed vertical lines),  
whereas the asymmetry of the SW profiles in the structure with nonreciprocal dispersion 
leads to the FMR excitation of every mode with $k=2 n \pi /a$ (in each of the bands shown in Fig.~\ref{disp}).


The positions of the peaks and their separation in the FMR spectrum of an MC with nonreciprocity 
can be estimated using the analytical formula for the dispersion relation in a uniform film 
with a periodic perturbation small enough to be assumed not to affect the frequency of the resonant modes 
(when band gaps open far from the BZ center). 
Since the asymmetry of the dispersion relation is only due to the DMI term in Eq.~(\ref{eq_1}), 
the separation between the peaks originating in modes with $k = \pm 2 \pi /a$ is:
\begin{equation}
\Delta f \approx \frac{4 \gamma D}{a M_{\text{S}}}\label{Eq:Delta_f}.
\end{equation}
Thus, the peak separation is proportional to $D$ and inversely proportional to $M_{S}$ and $a$, 
a property advantageous for experimental determination of the value of $D$. 
For the value used in this study, $D=1$ mJ/m$^2$, the calculated separation is 8.2~GHz. 
This can be compared with the numerical result, 6.9~GHz. 
The difference is due to too strong perturbation in the considered MC, the thickness of which changes between 1 and 0.5~nm.

In order to estimate the peak separation when the frequency is fixed and $H_{0}$ is varied, 
which is a common practice in FMR measurements, we derive the following formula for $\Delta H_{0}$:
\begin{widetext}
\begin{multline}
\mu_{0} \Delta H_{0} = \frac{2}{\gamma M_{\text{S}}}  \left[\sqrt{\left(f M_{\text{S}} \pi - \frac{2 D \gamma \pi}{a}\right)^2 + 
     e^{\frac{-4 d \pi}{a}} \left(\frac{M_{\text{S}} \omega_{M}}{4}\right)^2} - 
   \sqrt{ \left(f M_{\text{S}} \pi + \frac{2 D \gamma \pi}{a}\right)^2 + 
     e^{\frac{-4 d \pi}{a}} \left(\frac{M_{\text{S}} \omega_{M}}{4}\right)^2}\right].
\end{multline}
\end{widetext}
This complicated relation involves also the thickness~$d$. 
However, if $d/a \ll 1$, the $d$~dependence can be omitted:
\begin{eqnarray}
\mu_{0} \Delta H_{0} &\approx &
 \frac{2}{\gamma M_{S}}  \left[\sqrt{ \left(f M_{S} \pi - \frac{2 D \gamma \pi}{a}\right)^2 + 
      \left(\frac{M_{S} \omega_{M}}{4}\right)^2}\right. \nonumber \\
&-& \left.\sqrt{\left(f M_{S} \pi + \frac{2 D \gamma \pi}{a}\right)^2 + 
      \left(\frac{M_{S} \omega_{M}}{4}\right)^2}\right]\label{Eq:Delta_H}.
\end{eqnarray}
For the assumed parameter value $D=1$~mJ/m$^2$ and the frequency set to 16~GHz 
the value of $\mu_{0} \Delta H_{0}$ is around 175~mT. 
This field separation should be compared with the peak broadening due to damping. 
The width (full width at half maximum, FWHM) of a resonant peak 
is proportional to the Gilbert damping parameter and the frequency:\cite{berger2014magnetization}
\begin{equation}
\mu_{0} \Delta H_{\text{FWHM}}=\frac{4 \pi \alpha f}{\gamma}.
\label{FWHM}
\end{equation}
For the values used here $ \mu_{0} \Delta H_{\text{FWHM}}$ in a Co ultrathin film is around 10~mT, 
much smaller than the peak separation resulting from the DMI. 
This means that by measuring the peak separation in the FMR spectrum 
versus either frequency or field in MCs
we can easily estimate the DMI strength from Eq.~(\ref{Eq:Delta_f}) or (\ref{Eq:Delta_H}), respectively.

However, it is worthy of notice that in the case of MCs with a very small perturbation 
the detection of peaks might require high-precision and low-noise FMR measurements. 
On the other hand, a large perturbation in an MC will further increase the peaks in the FMR spectrum.

\section{FMR spectra of isolated stripes}

\begin{figure}[!ht]
\includegraphics[width=0.45\textwidth]{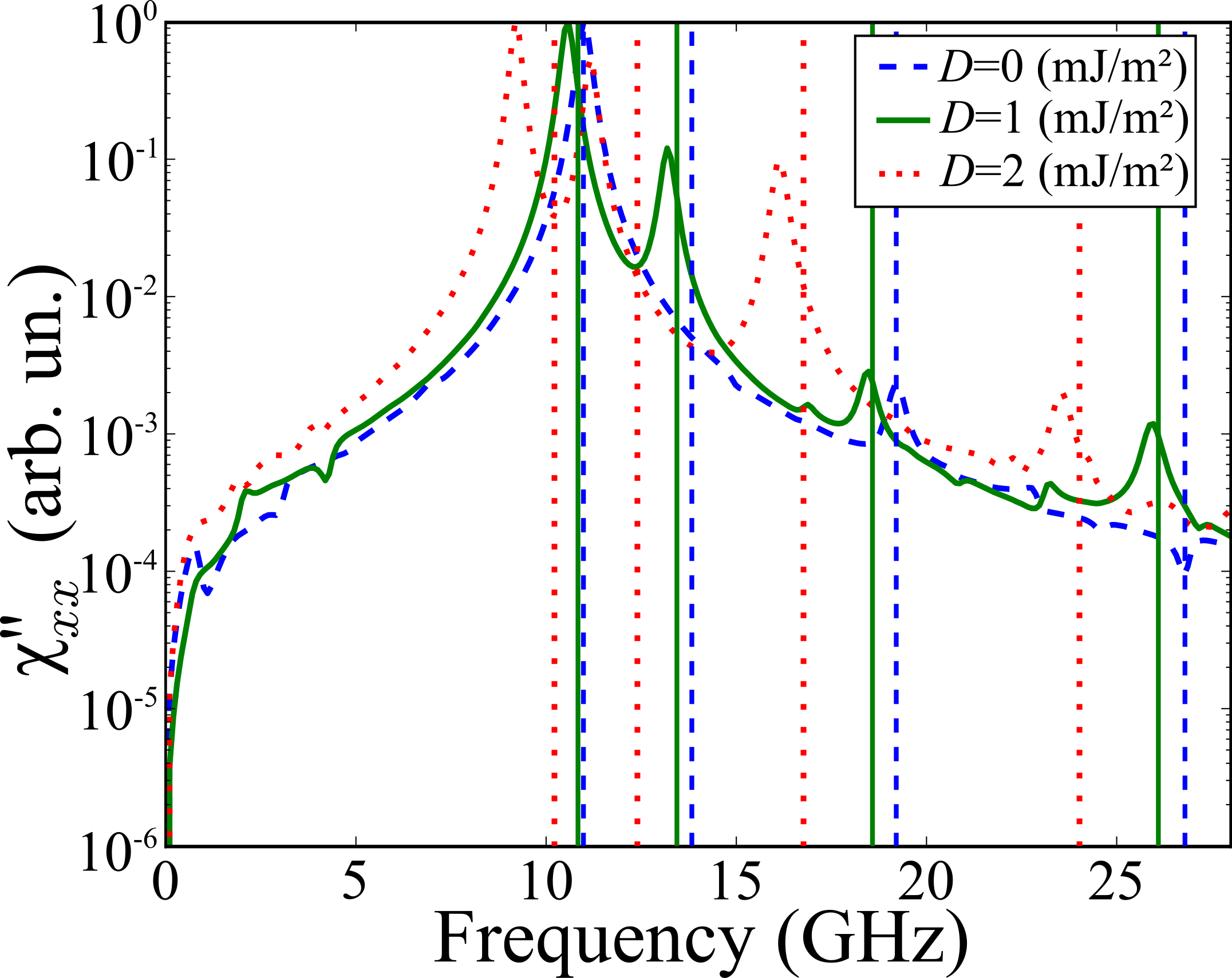}
\caption{(Color online) FMR frequency spectrum of a 1~nm~thick and 100~nm~wide isolated Co stripe: 
FDTD results for $D$~=~0~mJ/m$^2$ (blue dashed line), $D$~=~1~mJ/m$^2$ (green solid line) and $D$~=~2~mJ/m$^2$ (red dotted line), with a 0.01~T external magnetic field and $K_u = 0$. The  vertical lines indicate the solutions of of the FDFEM method.}
\label{FMR_stripe_a}
\end{figure}

In this Section we study the influence of the DMI on the FMR spectrum in 
an isolated stripe or an array of non-interacting stripes, 
i.e, stripes separated by a distance $w_{2} > w_{1}$ (see, Fig.~\ref{structure}(b)). 
Fig.~\ref{FMR_stripe_a} presents the FMR spectra obtained in FDTD simulations of a 1~nm~thick and 100~nm~wide isolated stripe. The calculated frequencies in FDFEM method are plotted as vertical lines in Fig. 5. As for weak DMI the agreement between FDFEM and FDTD is satisfactory, a frequency shift is observed when DMI is strong ($\approx$~2~mJ/m$^2$). It might be due to the assumptions used in the FDFEM method (linearization and collinear alignment of the magnetization). Also the mumax$^3$ has implemented the nontrivial DMI boundary conditions \cite{rohart2013skyrmion}, whereas in the FDFEM calculations the electromagnetic boundary conditions are fulfilled. \cite{Mruczkiewicz13} A pining of the magnetization at boundaries can appear in FDFEM solutions due to the dipolar pining. \cite{guslienko2002effective} The influence of these boundary conditions is weak on the static configuration. However, the DMI boundary conditions could influence on the dynamical magnetization components (thus also resonance frequencies) when DMI is high ($\approx$~2~mJ/m$^2$). As shown in Ref.~[\onlinecite{rohart2013skyrmion}], the bending of the magnetization is present in tangentially magnetized stripes. Thus, in our case the influence of the DMI boundary conditions is expected on  dynamical out-of-plane component, $m_{z}$.

\begin{figure}[!ht]
\includegraphics[width=0.45\textwidth]{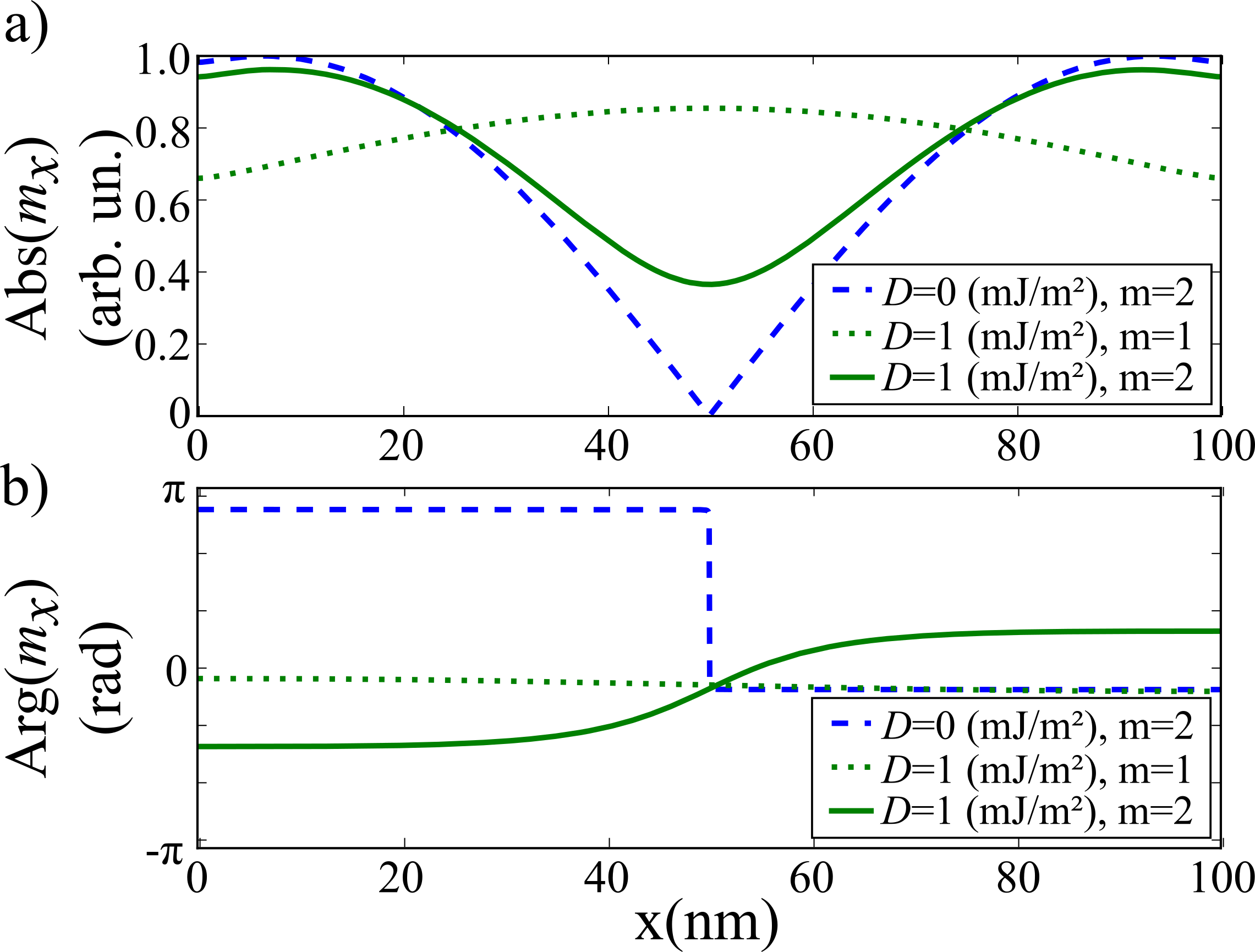}
\caption{(Color online)   (a) Amplitude and (b) phase profiles of SWs in a 100~nm wide stripe: 
2nd mode ($m = 2$, blue dashed line) with $D=0$~mJ/m$^2$, 
1st (fundamental) mode ($m = 1$, green dotted line) with $D=1$~mJ/m$^2$,
and 2nd mode ($m=2$, green continuous line) with $D=1$~mJ/m$^2$.}
\label{FMR_stripe_f}
\end{figure}

The confinement of the system results in the occurrence of standing modes, 
which can be excited with a uniform external microwave magnetic field. 
As in the case of MCs (discussed in Sec.~IV), the DMI is found to influence the FMR spectrum 
and increase the intensity of high-frequency excitations. 
The intensity increase is due to a significant modification of the mode profiles by the DMI, 
which changes the symmetries of the amplitude and phase distributions of the standing SWs. 
In the stripe without DMI the first standing SW mode above the fundamental excitation (Fig.~\ref{FMR_stripe_f}, blue dashed line) 
has an amplitude distribution
antisymmetric along the $x$~axis (magnetization oscillates in antiphase in the two halves of the stripe); 
as a result, the microwave field is not absorbed in FMR measurements. 
In the stripe with the DMI the phase of the amplitude 
of the second resonance mode (13.2~GHz) (Fig.~\ref{FMR_stripe_f}, green continuous line) 
changes continuously from $-\pi /2$ to $\pi /2$ along the $x$ direction. 
Thus, the amplitude of this mode does not have a node, 
which results in a relatively high intensity of the corresponding peak in the FMR spectrum. 

Interestingly, the DMI affects also the frequency and intensity of the fundamental excitation 
(mode 1 at 10.6~GHz) with a quasi-uniform amplitude (see, Fig.~\ref{FMR_stripe_f}). The frequency shift is due to a nonuniform amplitude distribution along the stripe width, 
resulting in an effective wavenumber along this direction 
and, consequently, a sensible influence of the DMI in accordance with Eq.~(\ref{eq_1}).
The change in the intensity is expected due to a slight phase shift between the oscillations of the magnetization 
and those of the excitation field (Fig.~\ref{FMR_stripe_f}(b), green dotted line). The change of the intensity is not visible in Fig.~\ref{FMR_stripe_a} because of the normalization to peak maximum, individual for each spectrum.
Since the intensity of higher-frequency resonance modes is proportional to the DMI strength, 
two peaks of similar intensity are observed in the structure with a strong DMI 
(see, Fig.~\ref{FMR_stripe_a} red dotted line, corresponding to $D=2$~mJ/m$^2$).


\begin{figure}[!ht]
\includegraphics[width=0.45\textwidth]{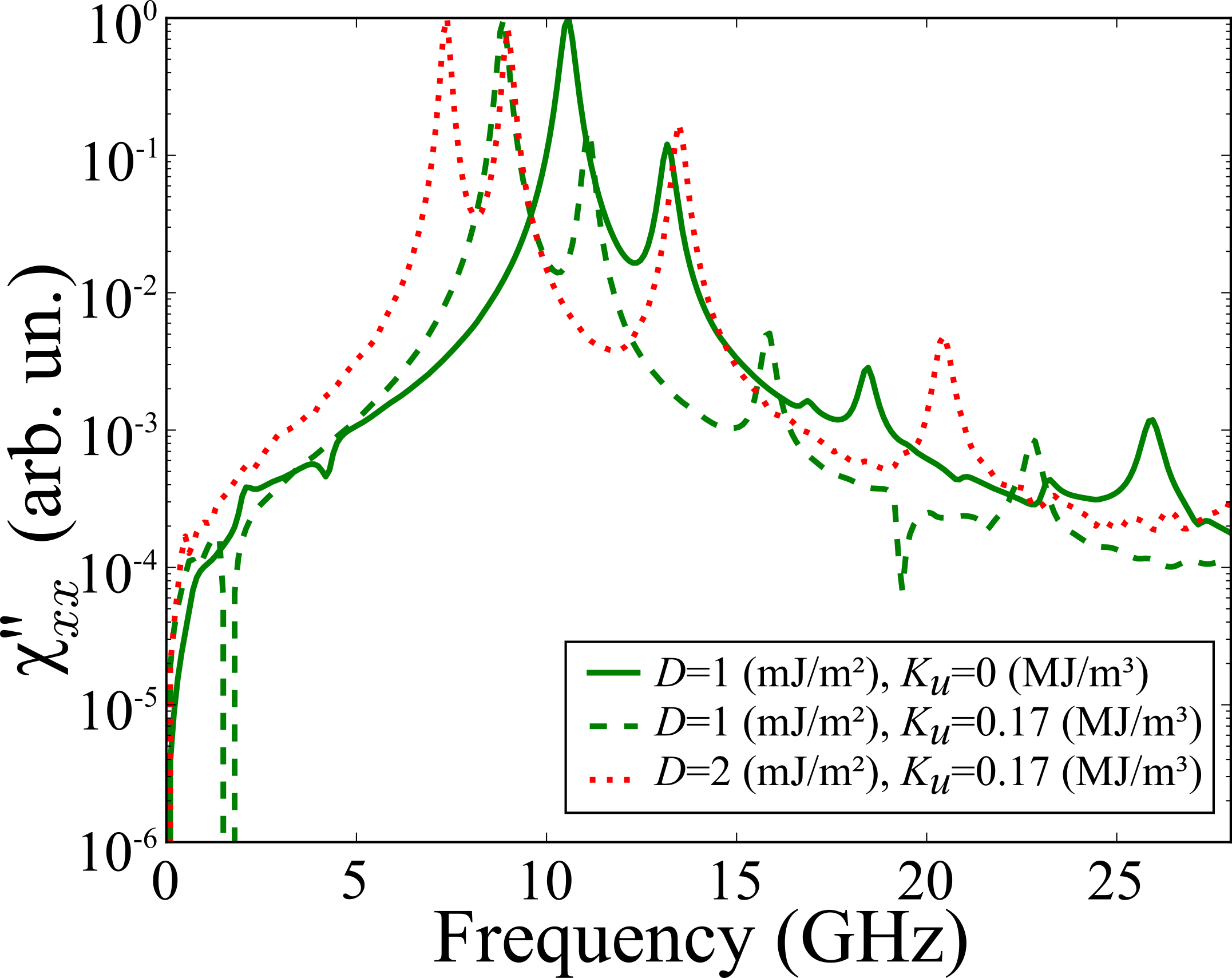}
\caption{(Color online) FMR frequency spectra of the isolated 1~nm~thick and 100~nm~wide Co~stripe with different magnetic anisotropy.  The results of the 
FDTD are shown for $D=1$~mJ/m$^2$, $K_{u}=0$~MJ/m$^3$ (green continuous line), 
$D=1$~mJ/m$^2$, $K_{u}=0.17$~MJ/m$^3$ (green dashed line), 
and $D=2$~mJ/m$^2$, $K_{u}=0.17$~MJ/m$^3$ (red dotted line).}
\label{FMR_stripe_b}
\end{figure}

Since the considered materials with DMI might also have a strong perpendicular uniaxial magnetic anisotropy, 
we have also studied its influence on the FMR spectrum with the FDTD method. 
Presented in Fig.~\ref{FMR_stripe_b} the results of our calculations 
performed with $K_{u}=0.17$~MJ/m$^3$ and two values of the DMI parameter, 
$D=1$ mJ/m$^2$ (green dashed line) and $D$~=~2~mJ/m$^2$ (red dotted line)
can be directly compared with the FMR spectrum 
obtained for $K_{u}=0$~MJ/m$^3$ and $D=1$~mJ/m$^2$ (green solid line, 
this is the same line as in Fig.~\ref{FMR_stripe_a}). 
In the case of nonzero~$K_u$ the intensities of the two low-frequency modes are preserved 
and equally shifted towards lower energies. 
The intensities of the higher-frequency modes show different behavior. 
For instance, the intensity of the 3rd mode increases, while the intensity of the 4th mode decreases with increasing $K_{u}$. However, 
the frequencies of all modes are reduced with increasing anisotropy.  

\begin{figure}[!ht]
\includegraphics[width=0.45\textwidth]{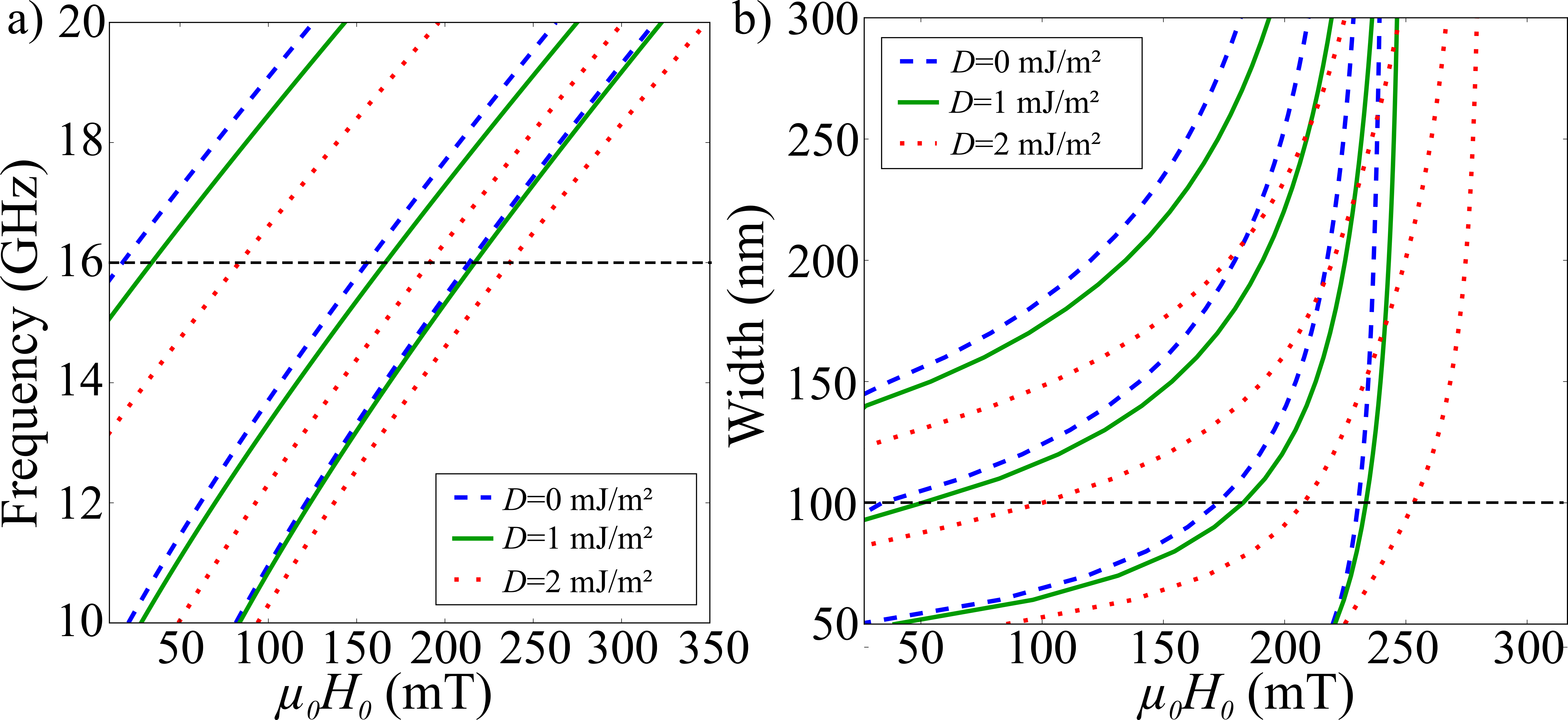}	
\caption{(Color online) FMR in a 1~nm~thick isolated Co stripe: 
(a) frequency vs. magnetic field dependence at fixed width, $w$~=~100~nm;
(b) stripe width vs. magnetic field dependence at a fixed frequency, $f$~=~16~GHz, 
for $D~=~0$~mJ/m$^2$ (blue dashed line), $D$~=~1~mJ/m$^2$ (green continuous line) and $D$~=~2~mJ/m$^2$ (red dotted line); 
$K_u$~=~0 in both plots. The horizontal black dashed lines indicate the frequency of stripe width corresponding to the FMR spectrum presented in Fig. \ref{FMR_stripe_field}. }
\label{FMR_stripe_5}
\end{figure}

Further calculations have been performed 
to determine the position of the peaks as a function of the stripe width and the magnetic field magnitude in the broad range of parameters for the selected values of $D$. The frequency-domain method was used here, due to its higher efficiency in calculations. Overestimated frequencies obtained from FDFEM for high values of the DMI do not influence qualitatively the obtained dependencies and conclusions. 
The dependencies $f(H_0)$ obtained for the four lowest-frequency modes are plotted in Fig.~\ref{FMR_stripe_5}(a). 
The presented frequency vs. magnetic field dependence
indicates that a frequency of ca. 16~GHz is sufficient to obtain at least three resonant peaks 
in a 100~nm wide stripe and $D=1$~mJ/m$^2$. With increasing $D$ this threshold frequency for estimation of DMI strength even decreases. Since the threshold of the DMI strength, defined by Eq.~(\ref{Eq:treshold}), is above the considered range of $D$ values, 
all magnetic  fields in the considered range will saturate the sample, 
and an optimum stripe width can be chosen based on the separation of the peaks. 

Fig.~\ref{FMR_stripe_5}(b) presents the stripe width vs. magnetic field dependence. It shows that if a stripe's width is of the order of hundreds of nanometers, the resolution can be sufficient to differentiate the resonance peaks (since $\mu_{0} \Delta H_{\text{FWHM}} \approx 10$~mT according to the Eq. (\ref{FWHM}) and estimate $D$. The separation between peaks increases with decreasing stripe width, however 
the number of observed modes will decrease, though,  for $100 <w < 150$~nm only 3 resonance lines will be present in FMR spectra. The figure shows also that 
the influence of $D$ on the resonance field of the fundamental mode depends on the width of the stripe.
The sensitivity to $D$ is slight in narrow stripes (see the lines around 200~mT for a 50~nm stripe) 
and increases with the stripe width. 
This dependence can be related to the magnetization pinning in homogeneous stripes. 
The pinning, which can result from the dipolar interaction,\cite{Guslienko2002} 
increases with the stripe width. 
The pinning at the edges of the stripe increases the effective wavenumber of the SW, 
as a result of which the influence of the DMI on the mode frequency is increased as well. 
Fig.~\ref{FMR_stripe_5}(a)
shows the SW frequency versus the external magnetic field for a fixed stripe width of 100~nm.

\begin{figure}[!ht]
\includegraphics[width=0.45\textwidth]{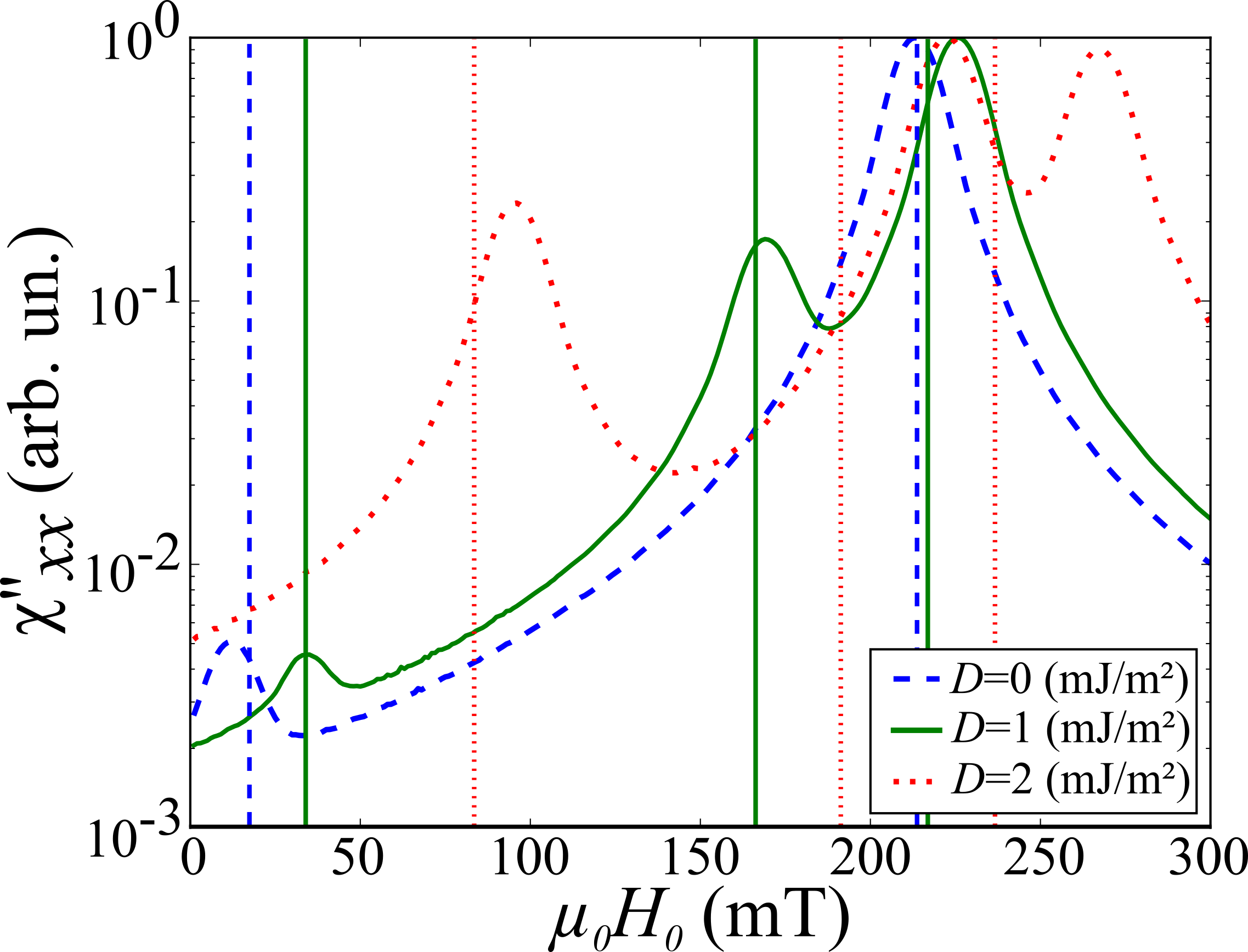}
\caption{(Color online) FMR external magnetic field spectrum of a 1~nm~thick and 100~nm~wide isolated Co stripe 
for $D$~=~0~mJ/m$^2$ (blue dashed line), $D$~=~1~mJ/m$^2$ (green solid line) and $D$~=~2~mJ/m$^2$ (red dotted line) at 16 GHz with $K_u = 0$ obtained from FDTD simulations. The positions of the FMR lines obtained with FDFEM are marked with vertical lines.}
\label{FMR_stripe_field}
\end{figure}

To validate the FDFEM results shown in Fig.~\ref{FMR_stripe_5} and numerically demonstrate the possibility of using magnetic field dependent FMR spectra for DMI estimation with the use of Eq.~(\ref{Eq:Delta_H}), we perform additional FDTD simulations.
 For the chosen frequency and the strip width, 16 GHz and 100 nm, respectively (parameters related to the horizontal black dashed lines in Fig. \ref{FMR_stripe_5}(a) and (b)) and three values of $D$, we plot numerical FMR external magnetic field spectrum in Fig.~\ref{FMR_stripe_field}. As expected, we found, that with increasing $D$ the resonance fields shift to higher fields and the excitations at low fields increase their FMR intensity. The Fig.~\ref{FMR_stripe_field} shows also that based on the FDFEM results presented in Fig. \ref{FMR_stripe_5} we are able to define a structure where FMR resonant peaks of quantized modes could be  intensive and differentiable in the external field spectrum. 

\section{Summary}\label{Sec:Conclusions}

The extended frequency-domain calculation model used in the paper 
shows a good agreement with micromagnetic simulations. It provides an efficient tool for fast characterization of magnonic structures 
with complex geometries, DMI and finite conductivities.

Using frequency domain method and micromagnetic simulations we have demonstrated a large impact of the DMI on the FMR spectrum 
and the profiles of quantized SW modes in 1D MCs and isolated stripe. 
In MCs the influence of the DMI on the FMR spectrum is due to the periodicity of the structure  
and the folding back of the magnonic bands to the 1st BZ. 
In stripes the impact of the DMI is related to the quantization of SWs due to confinement along the stripe width. 
We point out that the pinning of the magnetization at the stripe edges 
may have an important influence on the sensitivity of the fundamental mode to the DMI strength.

The findings presented in this paper provide the basis to propose 
 easy method for the determination of the DMI strength in ultrathin ferromagnetic films. This method use the dependence of the frequency separation between  the neighboring resonance peaks to the DMI strength. Moreover, we have derived an analytical approximate formula for the peak separation 
in the FMR field and frequency spectrum of small-perturbation MCs. It can be used for estimating the DMI strength and designing MCs with optimal band structure required for some applications.

The obtained results can also be of use for developing electromagnetic metamaterials 
proposed in Refs.~[\onlinecite{Mikhaylovskiy2010,PhysRevB.86.024425}], 
based on the interaction of the external microwave magnetic field with quantized SW modes. 
Providing another way of manipulating the FMR spectrum, especially increasing the intensity of the high frequency absorption peaks, 
the DMI can help to design materials with a negative refractive index in a broad and relatively high frequency range.  
The presented increase of coupling of quantized SWs with uniform external microwave magnetic field due to presence of the DMI can lead to search for analogous effect in other structures that support nonreciprocal propagation. \cite{verba2013conditions, Mruczkiewicz.2013b, lisenkov2015nonreciprocity, ma2014interfacial, di2015enhancement, liu2016nonreciprocal}

\section*{Acknowledgements}

The  project  is  financed  from  the SASPRO Programme.  The  research  leading  to  these results has received  funding from  the People Programme (Marie  Curie  Actions) European Union's Seventh Framework Programme under REA grant agreement No. 609427 (Project WEST: 1244/02/01). Research has been further co-funded by the Slovak Academy of Sciences and the European Union Horizon 2020 Research and Innovation Programme under Marie Sklodowska-Curie Grant Agreement No.~644348 (MagIC). The numerical calculations were performed at Poznan Supercomputing and Networking Center (Grant No.~209). 

\bibliographystyle{apsrev4-1}
\bibliography{books}

\begin{thebibliography}{38}%
\makeatletter
\providecommand \@ifxundefined [1]{%
 \@ifx{#1\undefined}
}%
\providecommand \@ifnum [1]{%
 \ifnum #1\expandafter \@firstoftwo
 \else \expandafter \@secondoftwo
 \fi
}%
\providecommand \@ifx [1]{%
 \ifx #1\expandafter \@firstoftwo
 \else \expandafter \@secondoftwo
 \fi
}%
\providecommand \natexlab [1]{#1}%
\providecommand \enquote  [1]{``#1''}%
\providecommand \bibnamefont  [1]{#1}%
\providecommand \bibfnamefont [1]{#1}%
\providecommand \citenamefont [1]{#1}%
\providecommand \href@noop [0]{\@secondoftwo}%
\providecommand \href [0]{\begingroup \@sanitize@url \@href}%
\providecommand \@href[1]{\@@startlink{#1}\@@href}%
\providecommand \@@href[1]{\endgroup#1\@@endlink}%
\providecommand \@sanitize@url [0]{\catcode `\\12\catcode `\$12\catcode
  `\&12\catcode `\#12\catcode `\^12\catcode `\_12\catcode `\%12\relax}%
\providecommand \@@startlink[1]{}%
\providecommand \@@endlink[0]{}%
\providecommand \url  [0]{\begingroup\@sanitize@url \@url }%
\providecommand \@url [1]{\endgroup\@href {#1}{\urlprefix }}%
\providecommand \urlprefix  [0]{URL }%
\providecommand \Eprint [0]{\href }%
\providecommand \doibase [0]{http://dx.doi.org/}%
\providecommand \selectlanguage [0]{\@gobble}%
\providecommand \bibinfo  [0]{\@secondoftwo}%
\providecommand \bibfield  [0]{\@secondoftwo}%
\providecommand \translation [1]{[#1]}%
\providecommand \BibitemOpen [0]{}%
\providecommand \bibitemStop [0]{}%
\providecommand \bibitemNoStop [0]{.\EOS\space}%
\providecommand \EOS [0]{\spacefactor3000\relax}%
\providecommand \BibitemShut  [1]{\csname bibitem#1\endcsname}%
\let\auto@bib@innerbib\@empty
\bibitem [{\citenamefont {Dzyaloshinsky}(1958)}]{117}%
  \BibitemOpen
  \bibfield  {author} {\bibinfo {author} {\bibfnamefont {I.}~\bibnamefont
  {Dzyaloshinsky}},\ }\href@noop {} {\bibfield  {journal} {\bibinfo  {journal}
  {Journal of Physics and Chemistry of Solids}\ }\textbf {\bibinfo {volume}
  {4}},\ \bibinfo {pages} {241} (\bibinfo {year} {1958})}\BibitemShut {NoStop}%
\bibitem [{\citenamefont {Moriya}(1960)}]{118}%
  \BibitemOpen
  \bibfield  {author} {\bibinfo {author} {\bibfnamefont {T.}~\bibnamefont
  {Moriya}},\ }\href@noop {} {\bibfield  {journal} {\bibinfo  {journal}
  {Physical Review}\ }\textbf {\bibinfo {volume} {120}},\ \bibinfo {pages} {91}
  (\bibinfo {year} {1960})}\BibitemShut {NoStop}%
\bibitem [{\citenamefont {Zhang}\ \emph {et~al.}(2015)\citenamefont {Zhang},
  \citenamefont {Zhao}, \citenamefont {Fangohr}, \citenamefont {Liu},
  \citenamefont {Xia}, \citenamefont {Xia},\ and\ \citenamefont
  {Morvan}}]{zhang2015skyrmion}%
  \BibitemOpen
  \bibfield  {author} {\bibinfo {author} {\bibfnamefont {X.}~\bibnamefont
  {Zhang}}, \bibinfo {author} {\bibfnamefont {G.}~\bibnamefont {Zhao}},
  \bibinfo {author} {\bibfnamefont {H.}~\bibnamefont {Fangohr}}, \bibinfo
  {author} {\bibfnamefont {J.~P.}\ \bibnamefont {Liu}}, \bibinfo {author}
  {\bibfnamefont {W.}~\bibnamefont {Xia}}, \bibinfo {author} {\bibfnamefont
  {J.}~\bibnamefont {Xia}}, \ and\ \bibinfo {author} {\bibfnamefont
  {F.}~\bibnamefont {Morvan}},\ }\href@noop {} {\bibfield  {journal} {\bibinfo
  {journal} {Scientific Reports}\ }\textbf {\bibinfo {volume} {5}},\ \bibinfo
  {pages} {7643} (\bibinfo {year} {2015})}\BibitemShut {NoStop}%
\bibitem [{\citenamefont {Stashkevich}\ \emph {et~al.}(2015)\citenamefont
  {Stashkevich}, \citenamefont {Belmeguenai}, \citenamefont {Roussign{\'e}},
  \citenamefont {Cherif}, \citenamefont {Kostylev}, \citenamefont {Gabor},
  \citenamefont {Lacour}, \citenamefont {Tiusan},\ and\ \citenamefont
  {Hehn}}]{stashkevich2015}%
  \BibitemOpen
  \bibfield  {author} {\bibinfo {author} {\bibfnamefont {A.~A.}\ \bibnamefont
  {Stashkevich}}, \bibinfo {author} {\bibfnamefont {M.}~\bibnamefont
  {Belmeguenai}}, \bibinfo {author} {\bibfnamefont {Y.}~\bibnamefont
  {Roussign{\'e}}}, \bibinfo {author} {\bibfnamefont {S.~M.}\ \bibnamefont
  {Cherif}}, \bibinfo {author} {\bibfnamefont {M.}~\bibnamefont {Kostylev}},
  \bibinfo {author} {\bibfnamefont {M.}~\bibnamefont {Gabor}}, \bibinfo
  {author} {\bibfnamefont {D.}~\bibnamefont {Lacour}}, \bibinfo {author}
  {\bibfnamefont {C.}~\bibnamefont {Tiusan}}, \ and\ \bibinfo {author}
  {\bibfnamefont {M.}~\bibnamefont {Hehn}},\ }\href@noop {} {\bibfield
  {journal} {\bibinfo  {journal} {Physical Review B}\ }\textbf {\bibinfo
  {volume} {91}},\ \bibinfo {pages} {214409} (\bibinfo {year}
  {2015})}\BibitemShut {NoStop}%
\bibitem [{\citenamefont {Boulle}\ \emph {et~al.}(2016)\citenamefont {Boulle},
  \citenamefont {Vogel}, \citenamefont {Yang}, \citenamefont {Pizzini},
  \citenamefont {Chaves}, \citenamefont {Locatelli}, \citenamefont {Sala},
  \citenamefont {Buda-Prejbeanu}, \citenamefont {Klein}, \citenamefont
  {Belmeguenai} \emph {et~al.}}]{boulle2016room}%
  \BibitemOpen
  \bibfield  {author} {\bibinfo {author} {\bibfnamefont {O.}~\bibnamefont
  {Boulle}}, \bibinfo {author} {\bibfnamefont {J.}~\bibnamefont {Vogel}},
  \bibinfo {author} {\bibfnamefont {H.}~\bibnamefont {Yang}}, \bibinfo {author}
  {\bibfnamefont {S.}~\bibnamefont {Pizzini}}, \bibinfo {author} {\bibfnamefont
  {D.~d.~S.}\ \bibnamefont {Chaves}}, \bibinfo {author} {\bibfnamefont
  {A.}~\bibnamefont {Locatelli}}, \bibinfo {author} {\bibfnamefont {T.~O.
  M.~A.}\ \bibnamefont {Sala}}, \bibinfo {author} {\bibfnamefont {L.~D.}\
  \bibnamefont {Buda-Prejbeanu}}, \bibinfo {author} {\bibfnamefont
  {O.}~\bibnamefont {Klein}}, \bibinfo {author} {\bibfnamefont
  {M.}~\bibnamefont {Belmeguenai}},  \emph {et~al.},\ }\href@noop {} {\bibfield
   {journal} {\bibinfo  {journal} {arXiv preprint arXiv:1601.02278}\ }
  (\bibinfo {year} {2016})}\BibitemShut {NoStop}%
\bibitem [{\citenamefont {Belmeguenai}\ \emph {et~al.}(2015)\citenamefont
  {Belmeguenai}, \citenamefont {Adam}, \citenamefont {Roussign{\'e}},
  \citenamefont {Eimer}, \citenamefont {Devolder}, \citenamefont {Kim},
  \citenamefont {Cherif}, \citenamefont {Stashkevich},\ and\ \citenamefont
  {Thiaville}}]{belmeguenai2015interfacial}%
  \BibitemOpen
  \bibfield  {author} {\bibinfo {author} {\bibfnamefont {M.}~\bibnamefont
  {Belmeguenai}}, \bibinfo {author} {\bibfnamefont {J.-P.}\ \bibnamefont
  {Adam}}, \bibinfo {author} {\bibfnamefont {Y.}~\bibnamefont {Roussign{\'e}}},
  \bibinfo {author} {\bibfnamefont {S.}~\bibnamefont {Eimer}}, \bibinfo
  {author} {\bibfnamefont {T.}~\bibnamefont {Devolder}}, \bibinfo {author}
  {\bibfnamefont {J.-V.}\ \bibnamefont {Kim}}, \bibinfo {author} {\bibfnamefont
  {S.~M.}\ \bibnamefont {Cherif}}, \bibinfo {author} {\bibfnamefont
  {A.}~\bibnamefont {Stashkevich}}, \ and\ \bibinfo {author} {\bibfnamefont
  {A.}~\bibnamefont {Thiaville}},\ }\href@noop {} {\bibfield  {journal}
  {\bibinfo  {journal} {Physical Review B}\ }\textbf {\bibinfo {volume} {91}},\
  \bibinfo {pages} {180405} (\bibinfo {year} {2015})}\BibitemShut {NoStop}%
\bibitem [{\citenamefont {K{\"o}rner}\ \emph {et~al.}(2015)\citenamefont
  {K{\"o}rner}, \citenamefont {Stigloher}, \citenamefont {Bauer}, \citenamefont
  {Hata}, \citenamefont {Taniguchi}, \citenamefont {Moriyama}, \citenamefont
  {Ono},\ and\ \citenamefont {Back}}]{korner2015interfacial}%
  \BibitemOpen
  \bibfield  {author} {\bibinfo {author} {\bibfnamefont {H.}~\bibnamefont
  {K{\"o}rner}}, \bibinfo {author} {\bibfnamefont {J.}~\bibnamefont
  {Stigloher}}, \bibinfo {author} {\bibfnamefont {H.}~\bibnamefont {Bauer}},
  \bibinfo {author} {\bibfnamefont {H.}~\bibnamefont {Hata}}, \bibinfo {author}
  {\bibfnamefont {T.}~\bibnamefont {Taniguchi}}, \bibinfo {author}
  {\bibfnamefont {T.}~\bibnamefont {Moriyama}}, \bibinfo {author}
  {\bibfnamefont {T.}~\bibnamefont {Ono}}, \ and\ \bibinfo {author}
  {\bibfnamefont {C.}~\bibnamefont {Back}},\ }\href@noop {} {\bibfield
  {journal} {\bibinfo  {journal} {Physical Review B}\ }\textbf {\bibinfo
  {volume} {92}},\ \bibinfo {pages} {220413} (\bibinfo {year}
  {2015})}\BibitemShut {NoStop}%
\bibitem [{\citenamefont {Lee}\ \emph {et~al.}(2015)\citenamefont {Lee},
  \citenamefont {Jang}, \citenamefont {Min}, \citenamefont {Lee}, \citenamefont
  {Lee},\ and\ \citenamefont {Chang}}]{lee2015all}%
  \BibitemOpen
  \bibfield  {author} {\bibinfo {author} {\bibfnamefont {J.~M.}\ \bibnamefont
  {Lee}}, \bibinfo {author} {\bibfnamefont {C.}~\bibnamefont {Jang}}, \bibinfo
  {author} {\bibfnamefont {B.-C.}\ \bibnamefont {Min}}, \bibinfo {author}
  {\bibfnamefont {S.-W.}\ \bibnamefont {Lee}}, \bibinfo {author} {\bibfnamefont
  {K.-J.}\ \bibnamefont {Lee}}, \ and\ \bibinfo {author} {\bibfnamefont
  {J.}~\bibnamefont {Chang}},\ }\href@noop {} {\bibfield  {journal} {\bibinfo
  {journal} {Nano letters}\ }\textbf {\bibinfo {volume} {16}},\ \bibinfo
  {pages} {62} (\bibinfo {year} {2015})}\BibitemShut {NoStop}%
\bibitem [{\citenamefont {Udvardi}\ and\ \citenamefont
  {Szunyogh}(2009)}]{udvardi2009chiral}%
  \BibitemOpen
  \bibfield  {author} {\bibinfo {author} {\bibfnamefont {L.}~\bibnamefont
  {Udvardi}}\ and\ \bibinfo {author} {\bibfnamefont {L.}~\bibnamefont
  {Szunyogh}},\ }\href@noop {} {\bibfield  {journal} {\bibinfo  {journal}
  {Physical Review Letters}\ }\textbf {\bibinfo {volume} {102}},\ \bibinfo
  {pages} {207204} (\bibinfo {year} {2009})}\BibitemShut {NoStop}%
\bibitem [{\citenamefont {Moon}\ \emph {et~al.}(2013)\citenamefont {Moon},
  \citenamefont {Seo}, \citenamefont {Lee}, \citenamefont {Kim}, \citenamefont
  {Ryu}, \citenamefont {Lee}, \citenamefont {McMichael},\ and\ \citenamefont
  {Stiles}}]{45}%
  \BibitemOpen
  \bibfield  {author} {\bibinfo {author} {\bibfnamefont {J.~H.}\ \bibnamefont
  {Moon}}, \bibinfo {author} {\bibfnamefont {S.~M.}\ \bibnamefont {Seo}},
  \bibinfo {author} {\bibfnamefont {K.~J.}\ \bibnamefont {Lee}}, \bibinfo
  {author} {\bibfnamefont {K.~W.}\ \bibnamefont {Kim}}, \bibinfo {author}
  {\bibfnamefont {J.}~\bibnamefont {Ryu}}, \bibinfo {author} {\bibfnamefont
  {H.~W.}\ \bibnamefont {Lee}}, \bibinfo {author} {\bibfnamefont {R.~D.}\
  \bibnamefont {McMichael}}, \ and\ \bibinfo {author} {\bibfnamefont {M.~D.}\
  \bibnamefont {Stiles}},\ }\href@noop {} {\bibfield  {journal} {\bibinfo
  {journal} {Physical Review B}\ }\textbf {\bibinfo {volume} {88}},\ \bibinfo
  {pages} {184404} (\bibinfo {year} {2013})}\BibitemShut {NoStop}%
\bibitem [{\citenamefont {Cort{\'e}s-Ortu{\~n}o}\ and\ \citenamefont
  {Landeros}(2013)}]{cortes2013influence}%
  \BibitemOpen
  \bibfield  {author} {\bibinfo {author} {\bibfnamefont {D.}~\bibnamefont
  {Cort{\'e}s-Ortu{\~n}o}}\ and\ \bibinfo {author} {\bibfnamefont
  {P.}~\bibnamefont {Landeros}},\ }\href@noop {} {\bibfield  {journal}
  {\bibinfo  {journal} {Journal of Physics: Condensed Matter}\ }\textbf
  {\bibinfo {volume} {25}},\ \bibinfo {pages} {156001} (\bibinfo {year}
  {2013})}\BibitemShut {NoStop}%
\bibitem [{\citenamefont {Ma}\ and\ \citenamefont
  {Zhou}(2014)}]{ma2014interfacial}%
  \BibitemOpen
  \bibfield  {author} {\bibinfo {author} {\bibfnamefont {F.}~\bibnamefont
  {Ma}}\ and\ \bibinfo {author} {\bibfnamefont {Y.}~\bibnamefont {Zhou}},\
  }\href@noop {} {\bibfield  {journal} {\bibinfo  {journal} {RSC Advances}\
  }\textbf {\bibinfo {volume} {4}},\ \bibinfo {pages} {46454} (\bibinfo {year}
  {2014})}\BibitemShut {NoStop}%
\bibitem [{\citenamefont {Mikhaylovskiy}\ \emph {et~al.}(2010)\citenamefont
  {Mikhaylovskiy}, \citenamefont {Hendry},\ and\ \citenamefont
  {Kruglyak}}]{Mikhaylovskiy2010}%
  \BibitemOpen
  \bibfield  {author} {\bibinfo {author} {\bibfnamefont {R.~V.}\ \bibnamefont
  {Mikhaylovskiy}}, \bibinfo {author} {\bibfnamefont {E.}~\bibnamefont
  {Hendry}}, \ and\ \bibinfo {author} {\bibfnamefont {V.~V.}\ \bibnamefont
  {Kruglyak}},\ }\href {\doibase 10.1103/PhysRevB.82.195446} {\bibfield
  {journal} {\bibinfo  {journal} {Phys. Rev. B}\ }\textbf {\bibinfo {volume}
  {82}},\ \bibinfo {pages} {195446} (\bibinfo {year} {2010})}\BibitemShut
  {NoStop}%
\bibitem [{\citenamefont {Mruczkiewicz}\ \emph {et~al.}(2012)\citenamefont
  {Mruczkiewicz}, \citenamefont {Krawczyk}, \citenamefont {Mikhaylovskiy},\
  and\ \citenamefont {Kruglyak}}]{PhysRevB.86.024425}%
  \BibitemOpen
  \bibfield  {author} {\bibinfo {author} {\bibfnamefont {M.}~\bibnamefont
  {Mruczkiewicz}}, \bibinfo {author} {\bibfnamefont {M.}~\bibnamefont
  {Krawczyk}}, \bibinfo {author} {\bibfnamefont {R.~V.}\ \bibnamefont
  {Mikhaylovskiy}}, \ and\ \bibinfo {author} {\bibfnamefont {V.~V.}\
  \bibnamefont {Kruglyak}},\ }\href {\doibase 10.1103/PhysRevB.86.024425}
  {\bibfield  {journal} {\bibinfo  {journal} {Phys. Rev. B}\ }\textbf {\bibinfo
  {volume} {86}},\ \bibinfo {pages} {024425} (\bibinfo {year}
  {2012})}\BibitemShut {NoStop}%
\bibitem [{\citenamefont {Mruczkiewicz}\ and\ \citenamefont
  {Krawczyk}(2014)}]{Mru}%
  \BibitemOpen
  \bibfield  {author} {\bibinfo {author} {\bibfnamefont {M.}~\bibnamefont
  {Mruczkiewicz}}\ and\ \bibinfo {author} {\bibfnamefont {M.}~\bibnamefont
  {Krawczyk}},\ }\href {\doibase http://dx.doi.org/10.1063/1.4868905}
  {\bibfield  {journal} {\bibinfo  {journal} {J. Appl. Phys.}\ }\textbf
  {\bibinfo {volume} {115}},\ \bibinfo {eid} {113909} (\bibinfo {year}
  {2014})}\BibitemShut {NoStop}%
\bibitem [{Com()}]{Comsol}%
  \BibitemOpen
  \href@noop {} {\bibinfo  {journal} {COMSOL Multiphysics® v. 5.2.
  www.comsol.com. COMSOL AB, Stockholm, Sweden.}\ }\BibitemShut {NoStop}%
\bibitem [{Note1()}]{Note1}%
  \BibitemOpen
\bibfield  {journal} {  }\bibinfo {note} {For instance FEniCS\cite
  {logg2012automated} or Agros2D\cite {karban2013numerical}. In particular, the
  functionality of FEniCS allows to define arbitrary equations, periodic
  boundary conditions and use of the eigenvalue solver, necessary for solving
  problem described in this manuscript. In addition, stated problem could be
  solved also with spectral methods, such as plane wave method implemented for
  SWs.\cite {krawczyk2008plane}}\BibitemShut {NoStop}%
\bibitem [{\citenamefont {Vansteenkiste}\ \emph {et~al.}(2014)\citenamefont
  {Vansteenkiste}, \citenamefont {Leliaert}, \citenamefont {Dvornik},
  \citenamefont {Helsen}, \citenamefont {Garcia-Sanchez},\ and\ \citenamefont
  {Van~Waeyenberge}}]{4899186}%
  \BibitemOpen
  \bibfield  {author} {\bibinfo {author} {\bibfnamefont {A.}~\bibnamefont
  {Vansteenkiste}}, \bibinfo {author} {\bibfnamefont {J.}~\bibnamefont
  {Leliaert}}, \bibinfo {author} {\bibfnamefont {M.}~\bibnamefont {Dvornik}},
  \bibinfo {author} {\bibfnamefont {M.}~\bibnamefont {Helsen}}, \bibinfo
  {author} {\bibfnamefont {F.}~\bibnamefont {Garcia-Sanchez}}, \ and\ \bibinfo
  {author} {\bibfnamefont {B.}~\bibnamefont {Van~Waeyenberge}},\ }\href
  {\doibase http://dx.doi.org/10.1063/1.4899186} {\bibfield  {journal}
  {\bibinfo  {journal} {AIP Advances}\ }\textbf {\bibinfo {volume} {4}},\
  \bibinfo {eid} {107133} (\bibinfo {year} {2014})}\BibitemShut {NoStop}%
\bibitem [{\citenamefont {Moreau-Luchaire}\ \emph {et~al.}(2016)\citenamefont
  {Moreau-Luchaire}, \citenamefont {Moutafis}, \citenamefont {Reyren},
  \citenamefont {Sampaio}, \citenamefont {Vaz}, \citenamefont {Van~Horne},
  \citenamefont {Bouzehouane}, \citenamefont {Garcia}, \citenamefont
  {Deranlot}, \citenamefont {Warnicke} \emph {et~al.}}]{moreau2016additive}%
  \BibitemOpen
  \bibfield  {author} {\bibinfo {author} {\bibfnamefont {C.}~\bibnamefont
  {Moreau-Luchaire}}, \bibinfo {author} {\bibfnamefont {C.}~\bibnamefont
  {Moutafis}}, \bibinfo {author} {\bibfnamefont {N.}~\bibnamefont {Reyren}},
  \bibinfo {author} {\bibfnamefont {J.}~\bibnamefont {Sampaio}}, \bibinfo
  {author} {\bibfnamefont {C.}~\bibnamefont {Vaz}}, \bibinfo {author}
  {\bibfnamefont {N.}~\bibnamefont {Van~Horne}}, \bibinfo {author}
  {\bibfnamefont {K.}~\bibnamefont {Bouzehouane}}, \bibinfo {author}
  {\bibfnamefont {K.}~\bibnamefont {Garcia}}, \bibinfo {author} {\bibfnamefont
  {C.}~\bibnamefont {Deranlot}}, \bibinfo {author} {\bibfnamefont
  {P.}~\bibnamefont {Warnicke}},  \emph {et~al.},\ }\href@noop {} {\bibfield
  {journal} {\bibinfo  {journal} {Nature Nanotechnology}\ }\textbf {\bibinfo
  {volume} {11}},\ \bibinfo {pages} {444} (\bibinfo {year} {2016})}\BibitemShut
  {NoStop}%
\bibitem [{\citenamefont {Beaujour}\ \emph {et~al.}(2006)\citenamefont
  {Beaujour}, \citenamefont {Lee}, \citenamefont {Kent}, \citenamefont
  {Krycka},\ and\ \citenamefont {Kao}}]{beaujour2006magnetization}%
  \BibitemOpen
  \bibfield  {author} {\bibinfo {author} {\bibfnamefont {J.~L.}\ \bibnamefont
  {Beaujour}}, \bibinfo {author} {\bibfnamefont {J.}~\bibnamefont {Lee}},
  \bibinfo {author} {\bibfnamefont {A.}~\bibnamefont {Kent}}, \bibinfo {author}
  {\bibfnamefont {K.}~\bibnamefont {Krycka}}, \ and\ \bibinfo {author}
  {\bibfnamefont {C.}~\bibnamefont {Kao}},\ }\href@noop {} {\bibfield
  {journal} {\bibinfo  {journal} {Physical Review B}\ }\textbf {\bibinfo
  {volume} {74}},\ \bibinfo {pages} {214405} (\bibinfo {year}
  {2006})}\BibitemShut {NoStop}%
\bibitem [{\citenamefont {Berger}\ \emph {et~al.}(2014)\citenamefont {Berger},
  \citenamefont {Amamou}, \citenamefont {White}, \citenamefont {Adur},
  \citenamefont {Pu}, \citenamefont {Kawakami},\ and\ \citenamefont
  {Hammel}}]{berger2014magnetization}%
  \BibitemOpen
  \bibfield  {author} {\bibinfo {author} {\bibfnamefont {A.}~\bibnamefont
  {Berger}}, \bibinfo {author} {\bibfnamefont {W.}~\bibnamefont {Amamou}},
  \bibinfo {author} {\bibfnamefont {S.}~\bibnamefont {White}}, \bibinfo
  {author} {\bibfnamefont {R.}~\bibnamefont {Adur}}, \bibinfo {author}
  {\bibfnamefont {Y.}~\bibnamefont {Pu}}, \bibinfo {author} {\bibfnamefont
  {R.}~\bibnamefont {Kawakami}}, \ and\ \bibinfo {author} {\bibfnamefont
  {P.~C.}\ \bibnamefont {Hammel}},\ }\href@noop {} {\bibfield  {journal}
  {\bibinfo  {journal} {Journal of Applied Physics}\ }\textbf {\bibinfo
  {volume} {115}},\ \bibinfo {pages} {17C510} (\bibinfo {year}
  {2014})}\BibitemShut {NoStop}%
\bibitem [{\citenamefont {Beginin}\ \emph {et~al.}(2012)\citenamefont
  {Beginin}, \citenamefont {Filimonov}, \citenamefont {Pavlov}, \citenamefont
  {Vysotskii},\ and\ \citenamefont {Nikitov}}]{beginin2012bragg}%
  \BibitemOpen
  \bibfield  {author} {\bibinfo {author} {\bibfnamefont {E.}~\bibnamefont
  {Beginin}}, \bibinfo {author} {\bibfnamefont {Y.~A.}\ \bibnamefont
  {Filimonov}}, \bibinfo {author} {\bibfnamefont {E.}~\bibnamefont {Pavlov}},
  \bibinfo {author} {\bibfnamefont {S.}~\bibnamefont {Vysotskii}}, \ and\
  \bibinfo {author} {\bibfnamefont {S.}~\bibnamefont {Nikitov}},\ }\href@noop
  {} {\bibfield  {journal} {\bibinfo  {journal} {Applied Physics Letters}\
  }\textbf {\bibinfo {volume} {100}},\ \bibinfo {pages} {252412} (\bibinfo
  {year} {2012})}\BibitemShut {NoStop}%
\bibitem [{\citenamefont {Mruczkiewicz}\ \emph
  {et~al.}(2013{\natexlab{a}})\citenamefont {Mruczkiewicz}, \citenamefont
  {Krawczyk}, \citenamefont {Sakharov}, \citenamefont {Khivintsev},
  \citenamefont {Filimonov},\ and\ \citenamefont {Nikitov}}]{Mruczkiewicz13}%
  \BibitemOpen
  \bibfield  {author} {\bibinfo {author} {\bibfnamefont {M.}~\bibnamefont
  {Mruczkiewicz}}, \bibinfo {author} {\bibfnamefont {M.}~\bibnamefont
  {Krawczyk}}, \bibinfo {author} {\bibfnamefont {V.~K.}\ \bibnamefont
  {Sakharov}}, \bibinfo {author} {\bibfnamefont {Y.~V.}\ \bibnamefont
  {Khivintsev}}, \bibinfo {author} {\bibfnamefont {Y.~A.}\ \bibnamefont
  {Filimonov}}, \ and\ \bibinfo {author} {\bibfnamefont {S.~A.}\ \bibnamefont
  {Nikitov}},\ }\href {\doibase 10.1063/1.4793085} {\bibfield  {journal}
  {\bibinfo  {journal} {J. Appl. Phys.}\ }\textbf {\bibinfo {volume} {113}},\
  \bibinfo {pages} {093908} (\bibinfo {year} {2013}{\natexlab{a}})}\BibitemShut
  {NoStop}%
\bibitem [{\citenamefont {Di}\ \emph {et~al.}(2015{\natexlab{a}})\citenamefont
  {Di}, \citenamefont {Zhang}, \citenamefont {Lim}, \citenamefont {Ng},
  \citenamefont {Kuok}, \citenamefont {Qiu},\ and\ \citenamefont
  {Yang}}]{di2015asymmetric}%
  \BibitemOpen
  \bibfield  {author} {\bibinfo {author} {\bibfnamefont {K.}~\bibnamefont
  {Di}}, \bibinfo {author} {\bibfnamefont {V.~L.}\ \bibnamefont {Zhang}},
  \bibinfo {author} {\bibfnamefont {H.~S.}\ \bibnamefont {Lim}}, \bibinfo
  {author} {\bibfnamefont {S.~C.}\ \bibnamefont {Ng}}, \bibinfo {author}
  {\bibfnamefont {M.~H.}\ \bibnamefont {Kuok}}, \bibinfo {author}
  {\bibfnamefont {X.}~\bibnamefont {Qiu}}, \ and\ \bibinfo {author}
  {\bibfnamefont {H.}~\bibnamefont {Yang}},\ }\href@noop {} {\bibfield
  {journal} {\bibinfo  {journal} {Applied Physics Letters}\ }\textbf {\bibinfo
  {volume} {106}},\ \bibinfo {pages} {052403} (\bibinfo {year}
  {2015}{\natexlab{a}})}\BibitemShut {NoStop}%
\bibitem [{\citenamefont {Yeh}(1979)}]{yeh1979electromagnetic}%
  \BibitemOpen
  \bibfield  {author} {\bibinfo {author} {\bibfnamefont {P.}~\bibnamefont
  {Yeh}},\ }\href@noop {} {\bibfield  {journal} {\bibinfo  {journal} {J. Opt.
  Soc. Am. A}\ }\textbf {\bibinfo {volume} {69}},\ \bibinfo {pages} {742}
  (\bibinfo {year} {1979})}\BibitemShut {NoStop}%
\bibitem [{\citenamefont {Mruczkiewicz}\ \emph
  {et~al.}(2013{\natexlab{b}})\citenamefont {Mruczkiewicz}, \citenamefont
  {Krawczyk}, \citenamefont {Gubbiotti}, \citenamefont {Tacchi}, \citenamefont
  {Filimonov}, \citenamefont {Kalyabin}, \citenamefont {Lisenkov},\ and\
  \citenamefont {Nikitov}}]{Mruczkiewicz.2013b}%
  \BibitemOpen
  \bibfield  {author} {\bibinfo {author} {\bibfnamefont {M.}~\bibnamefont
  {Mruczkiewicz}}, \bibinfo {author} {\bibfnamefont {M.}~\bibnamefont
  {Krawczyk}}, \bibinfo {author} {\bibfnamefont {G.}~\bibnamefont {Gubbiotti}},
  \bibinfo {author} {\bibfnamefont {S.}~\bibnamefont {Tacchi}}, \bibinfo
  {author} {\bibfnamefont {Y.~A.}\ \bibnamefont {Filimonov}}, \bibinfo {author}
  {\bibfnamefont {D.~V.}\ \bibnamefont {Kalyabin}}, \bibinfo {author}
  {\bibfnamefont {I.~V.}\ \bibnamefont {Lisenkov}}, \ and\ \bibinfo {author}
  {\bibfnamefont {S.~A.}\ \bibnamefont {Nikitov}},\ }\href@noop {} {\bibfield
  {journal} {\bibinfo  {journal} {New J. Phys.}\ }\textbf {\bibinfo {volume}
  {15}},\ \bibinfo {pages} {113023} (\bibinfo {year}
  {2013}{\natexlab{b}})}\BibitemShut {NoStop}%
\bibitem [{\citenamefont {Lisenkov}\ \emph {et~al.}(2015)\citenamefont
  {Lisenkov}, \citenamefont {Kalyabin}, \citenamefont {Osokin}, \citenamefont
  {Klos}, \citenamefont {Krawczyk},\ and\ \citenamefont
  {Nikitov}}]{lisenkov2015nonreciprocity}%
  \BibitemOpen
  \bibfield  {author} {\bibinfo {author} {\bibfnamefont {I.}~\bibnamefont
  {Lisenkov}}, \bibinfo {author} {\bibfnamefont {D.}~\bibnamefont {Kalyabin}},
  \bibinfo {author} {\bibfnamefont {S.}~\bibnamefont {Osokin}}, \bibinfo
  {author} {\bibfnamefont {J.}~\bibnamefont {Klos}}, \bibinfo {author}
  {\bibfnamefont {M.}~\bibnamefont {Krawczyk}}, \ and\ \bibinfo {author}
  {\bibfnamefont {S.}~\bibnamefont {Nikitov}},\ }\href@noop {} {\bibfield
  {journal} {\bibinfo  {journal} {Journal of Magnetism and Magnetic Materials}\
  }\textbf {\bibinfo {volume} {378}},\ \bibinfo {pages} {313} (\bibinfo {year}
  {2015})}\BibitemShut {NoStop}%
\bibitem [{\citenamefont {Bihler}\ \emph {et~al.}(2009)\citenamefont {Bihler},
  \citenamefont {Schoch}, \citenamefont {Limmer}, \citenamefont {Goennenwein},\
  and\ \citenamefont {Brandt}}]{bihler2009spin}%
  \BibitemOpen
  \bibfield  {author} {\bibinfo {author} {\bibfnamefont {C.}~\bibnamefont
  {Bihler}}, \bibinfo {author} {\bibfnamefont {W.}~\bibnamefont {Schoch}},
  \bibinfo {author} {\bibfnamefont {W.}~\bibnamefont {Limmer}}, \bibinfo
  {author} {\bibfnamefont {S.}~\bibnamefont {Goennenwein}}, \ and\ \bibinfo
  {author} {\bibfnamefont {M.}~\bibnamefont {Brandt}},\ }\href@noop {}
  {\bibfield  {journal} {\bibinfo  {journal} {Physical Review B}\ }\textbf
  {\bibinfo {volume} {79}},\ \bibinfo {pages} {045205} (\bibinfo {year}
  {2009})}\BibitemShut {NoStop}%
\bibitem [{\citenamefont {G{\'e}rardin}\ \emph {et~al.}(2000)\citenamefont
  {G{\'e}rardin}, \citenamefont {Youssef}, \citenamefont {Le~Gall},
  \citenamefont {Vukadinovic}, \citenamefont {Jacquart},\ and\ \citenamefont
  {Donahue}}]{gerardin2000micromagnetics}%
  \BibitemOpen
  \bibfield  {author} {\bibinfo {author} {\bibfnamefont {O.}~\bibnamefont
  {G{\'e}rardin}}, \bibinfo {author} {\bibfnamefont {J.~B.}\ \bibnamefont
  {Youssef}}, \bibinfo {author} {\bibfnamefont {H.}~\bibnamefont {Le~Gall}},
  \bibinfo {author} {\bibfnamefont {N.}~\bibnamefont {Vukadinovic}}, \bibinfo
  {author} {\bibfnamefont {P.}~\bibnamefont {Jacquart}}, \ and\ \bibinfo
  {author} {\bibfnamefont {M.}~\bibnamefont {Donahue}},\ }\href@noop {}
  {\bibfield  {journal} {\bibinfo  {journal} {Journal of Applied Physics}\
  }\textbf {\bibinfo {volume} {88}},\ \bibinfo {pages} {5899} (\bibinfo {year}
  {2000})}\BibitemShut {NoStop}%
\bibitem [{\citenamefont {Rohart}\ and\ \citenamefont
  {Thiaville}(2013)}]{rohart2013skyrmion}%
  \BibitemOpen
  \bibfield  {author} {\bibinfo {author} {\bibfnamefont {S.}~\bibnamefont
  {Rohart}}\ and\ \bibinfo {author} {\bibfnamefont {A.}~\bibnamefont
  {Thiaville}},\ }\href@noop {} {\bibfield  {journal} {\bibinfo  {journal}
  {Physical Review B}\ }\textbf {\bibinfo {volume} {88}},\ \bibinfo {pages}
  {184422} (\bibinfo {year} {2013})}\BibitemShut {NoStop}%
\bibitem [{\citenamefont {Guslienko}\ \emph
  {et~al.}(2002{\natexlab{a}})\citenamefont {Guslienko}, \citenamefont
  {Demokritov}, \citenamefont {Hillebrands},\ and\ \citenamefont
  {Slavin}}]{guslienko2002effective}%
  \BibitemOpen
  \bibfield  {author} {\bibinfo {author} {\bibfnamefont {K.~Y.}\ \bibnamefont
  {Guslienko}}, \bibinfo {author} {\bibfnamefont {S.}~\bibnamefont
  {Demokritov}}, \bibinfo {author} {\bibfnamefont {B.}~\bibnamefont
  {Hillebrands}}, \ and\ \bibinfo {author} {\bibfnamefont {A.}~\bibnamefont
  {Slavin}},\ }\href@noop {} {\bibfield  {journal} {\bibinfo  {journal}
  {Physical Review B}\ }\textbf {\bibinfo {volume} {66}},\ \bibinfo {pages}
  {132402} (\bibinfo {year} {2002}{\natexlab{a}})}\BibitemShut {NoStop}%
\bibitem [{\citenamefont {Guslienko}\ \emph
  {et~al.}(2002{\natexlab{b}})\citenamefont {Guslienko}, \citenamefont
  {Demokritov}, \citenamefont {Hillebrands},\ and\ \citenamefont
  {Slavin}}]{Guslienko2002}%
  \BibitemOpen
  \bibfield  {author} {\bibinfo {author} {\bibfnamefont {K.~Y.}\ \bibnamefont
  {Guslienko}}, \bibinfo {author} {\bibfnamefont {S.~O.}\ \bibnamefont
  {Demokritov}}, \bibinfo {author} {\bibfnamefont {B.}~\bibnamefont
  {Hillebrands}}, \ and\ \bibinfo {author} {\bibfnamefont {A.~N.}\ \bibnamefont
  {Slavin}},\ }\href {\doibase 10.1103/PhysRevB.66.132402} {\bibfield
  {journal} {\bibinfo  {journal} {Phys. Rev. B}\ }\textbf {\bibinfo {volume}
  {66}},\ \bibinfo {pages} {132402} (\bibinfo {year}
  {2002}{\natexlab{b}})}\BibitemShut {NoStop}%
\bibitem [{\citenamefont {Verba}\ \emph {et~al.}(2013)\citenamefont {Verba},
  \citenamefont {Tiberkevich}, \citenamefont {Bankowski}, \citenamefont
  {Meitzler}, \citenamefont {Melkov},\ and\ \citenamefont
  {Slavin}}]{verba2013conditions}%
  \BibitemOpen
  \bibfield  {author} {\bibinfo {author} {\bibfnamefont {R.}~\bibnamefont
  {Verba}}, \bibinfo {author} {\bibfnamefont {V.}~\bibnamefont {Tiberkevich}},
  \bibinfo {author} {\bibfnamefont {E.}~\bibnamefont {Bankowski}}, \bibinfo
  {author} {\bibfnamefont {T.}~\bibnamefont {Meitzler}}, \bibinfo {author}
  {\bibfnamefont {G.}~\bibnamefont {Melkov}}, \ and\ \bibinfo {author}
  {\bibfnamefont {A.}~\bibnamefont {Slavin}},\ }\href@noop {} {\bibfield
  {journal} {\bibinfo  {journal} {Applied Physics Letters}\ }\textbf {\bibinfo
  {volume} {103}},\ \bibinfo {pages} {082407} (\bibinfo {year}
  {2013})}\BibitemShut {NoStop}%
\bibitem [{\citenamefont {Di}\ \emph {et~al.}(2015{\natexlab{b}})\citenamefont
  {Di}, \citenamefont {Feng}, \citenamefont {Piramanayagam}, \citenamefont
  {Zhang}, \citenamefont {Lim}, \citenamefont {Ng},\ and\ \citenamefont
  {Kuok}}]{di2015enhancement}%
  \BibitemOpen
  \bibfield  {author} {\bibinfo {author} {\bibfnamefont {K.}~\bibnamefont
  {Di}}, \bibinfo {author} {\bibfnamefont {S.}~\bibnamefont {Feng}}, \bibinfo
  {author} {\bibfnamefont {S.}~\bibnamefont {Piramanayagam}}, \bibinfo {author}
  {\bibfnamefont {V.}~\bibnamefont {Zhang}}, \bibinfo {author} {\bibfnamefont
  {H.}~\bibnamefont {Lim}}, \bibinfo {author} {\bibfnamefont {S.}~\bibnamefont
  {Ng}}, \ and\ \bibinfo {author} {\bibfnamefont {M.}~\bibnamefont {Kuok}},\
  }\href@noop {} {\bibfield  {journal} {\bibinfo  {journal} {Scientific
  Reports}\ }\textbf {\bibinfo {volume} {5}},\ \bibinfo {pages} {10153}
  (\bibinfo {year} {2015}{\natexlab{b}})}\BibitemShut {NoStop}%
\bibitem [{\citenamefont {Liu}\ \emph {et~al.}(2016)\citenamefont {Liu},
  \citenamefont {Shen}, \citenamefont {Fang},\ and\ \citenamefont
  {Jin}}]{liu2016nonreciprocal}%
  \BibitemOpen
  \bibfield  {author} {\bibinfo {author} {\bibfnamefont {W.}~\bibnamefont
  {Liu}}, \bibinfo {author} {\bibfnamefont {Y.}~\bibnamefont {Shen}}, \bibinfo
  {author} {\bibfnamefont {G.}~\bibnamefont {Fang}}, \ and\ \bibinfo {author}
  {\bibfnamefont {C.}~\bibnamefont {Jin}},\ }\href@noop {} {\bibfield
  {journal} {\bibinfo  {journal} {Journal of Physics: Condensed Matter}\
  }\textbf {\bibinfo {volume} {28}},\ \bibinfo {pages} {196001} (\bibinfo
  {year} {2016})}\BibitemShut {NoStop}%
\bibitem [{\citenamefont {Logg}\ \emph {et~al.}(2012)\citenamefont {Logg},
  \citenamefont {Mardal},\ and\ \citenamefont {Wells}}]{logg2012automated}%
  \BibitemOpen
  \bibfield  {author} {\bibinfo {author} {\bibfnamefont {A.}~\bibnamefont
  {Logg}}, \bibinfo {author} {\bibfnamefont {K.-A.}\ \bibnamefont {Mardal}}, \
  and\ \bibinfo {author} {\bibfnamefont {G.}~\bibnamefont {Wells}},\
  }\href@noop {} {\emph {\bibinfo {title} {Automated solution of differential
  equations by the finite element method: The FEniCS book}}},\ Vol.~\bibinfo
  {volume} {84}\ (\bibinfo  {publisher} {Springer Science \& Business Media},\
  \bibinfo {year} {2012})\BibitemShut {NoStop}%
\bibitem [{\citenamefont {Karban}\ \emph {et~al.}(2013)\citenamefont {Karban},
  \citenamefont {Mach}, \citenamefont {K{\o u}s}, \citenamefont {P{\'a}nek},\
  and\ \citenamefont {Dole{\v{z}}el}}]{karban2013numerical}%
  \BibitemOpen
  \bibfield  {author} {\bibinfo {author} {\bibfnamefont {P.}~\bibnamefont
  {Karban}}, \bibinfo {author} {\bibfnamefont {F.}~\bibnamefont {Mach}},
  \bibinfo {author} {\bibfnamefont {P.}~\bibnamefont {K{\o u}s}}, \bibinfo
  {author} {\bibfnamefont {D.}~\bibnamefont {P{\'a}nek}}, \ and\ \bibinfo
  {author} {\bibfnamefont {I.}~\bibnamefont {Dole{\v{z}}el}},\ }\href@noop {}
  {\bibfield  {journal} {\bibinfo  {journal} {Computing}\ }\textbf {\bibinfo
  {volume} {95}},\ \bibinfo {pages} {381} (\bibinfo {year} {2013})}\BibitemShut
  {NoStop}%
\bibitem [{\citenamefont {Krawczyk}\ and\ \citenamefont
  {Puszkarski}(2008)}]{krawczyk2008plane}%
  \BibitemOpen
  \bibfield  {author} {\bibinfo {author} {\bibfnamefont {M.}~\bibnamefont
  {Krawczyk}}\ and\ \bibinfo {author} {\bibfnamefont {H.}~\bibnamefont
  {Puszkarski}},\ }\href@noop {} {\bibfield  {journal} {\bibinfo  {journal}
  {Physical Review B}\ }\textbf {\bibinfo {volume} {77}},\ \bibinfo {pages}
  {054437} (\bibinfo {year} {2008})}\BibitemShut {NoStop}%
\end{thebibliography}%

\end{document}